%% file: tps_part_1.tex
\documentclass[journal]{IEEEtran}
\usepackage{cite}

\usepackage[cmex10]{amsmath}
\usepackage{amssymb,amsfonts}
\usepackage{algorithmic,algorithm}
\usepackage{amsthm}
\usepackage{caption}
\usepackage{bm}
\usepackage{placeins}

\usepackage{graphicx}
\usepackage{etoolbox}
\usepackage[caption=false,font=footnotesize]{subfig}
\usepackage{multirow}
\usepackage{pgfplots}
\usepackage{pgfplotstable}
\usepackage{comment}

\pgfplotsset{compat=newest,
    /pgfplots/ybar legend/.style={
    /pgfplots/legend image code/.code={%
       \draw[##1,/tikz/.cd,yshift=-0.25em]
        (0cm,0cm) rectangle (3pt,0.8em);},
   },
}

\newtoggle{isreport}
\toggletrue{isreport}

\newtoggle{isarxiv}
\toggletrue{isarxiv}

\newtheorem{lemma}{Lemma}

\newtheorem{prop}[lemma]{Proposition}
\newtheorem{cor}[lemma]{Corollary}
\newtheorem{thm}[lemma]{Theorem}
\theoremstyle{remark}
\theoremstyle{theorem}\newtheorem{eg}{Example}
\theoremstyle{theorem}\newtheorem{defn}[lemma]{Definition}

\newcommand{\paren}[1]{\left(#1\right)}

\newcommand{\set}[1]{\left\{#1\right\}}
\newcommand{\abs}[1]{\left|#1\right|}

\newcommand{\normm}[1]{{\left\vert\kern-0.25ex\left\vert\kern-0.25ex\left\vert #1
    \right\vert\kern-0.25ex\right\vert\kern-0.25ex\right\vert}}

\DeclareMathOperator{\spann}{span}

\DeclareMathOperator{\diag}{diag}

\newcommand{\R}{\mathbb R}

\newcommand{\ol}[1]{\overline{#1}}

\newcommand{\bs}{\backslash}

\newcommand{\calE}{\mathcal{E}}

\newcommand{\calG}{\mathcal{G}}

\newcommand{\calL}{\mathcal{L}}

\newcommand{\calN}{\mathcal{N}}

\newcommand{\calP}{\mathcal{P}}

\newcommand{\calT}{\mathcal{T}}

\newcommand{\bff}[1]{{\bf #1}}

\usepackage{color}
\usepackage{enumerate}

\definecolor{myred}{RGB}{202,0,32}
\definecolor{myorange}{RGB}{244,165,130}
\definecolor{myviolet}{RGB}{194,165,207}
\definecolor{mycyan}{RGB}{146,197,222}
\definecolor{myblue}{RGB}{5,113,176}
\definecolor{mygreen}{RGB}{127,191,123}
\definecolor{mytile}{RGB}{27,120,55}

\newcommand\slow[1]{\textcolor{myred}{[SL: #1]}}

\usepackage{tikz}
\newcommand*\circled[1]{\tikz[baseline=(char.base)]{
            \node[shape=circle,draw,inner sep=0.4pt] (char) {#1};}}

\IEEEoverridecommandlockouts

\title{
Line Failure Localization of Power Networks
\\ Part I: Non-cut Outages
}

\author{Linqi Guo, Chen~Liang, Alessandro~Zocca, Steven H.~Low,~and~Adam~Wierman
\thanks{This work has been supported by Resnick Fellowship, Linde Institute Research Award, NWO Rubicon grant 680.50.1529, 
	NSF through grants CCF 1637598, ECCS 1619352, ECCS 1931662, CNS 1545096, CNS 1518941, CPS ECCS 1739355, CPS 154471.}
\thanks{LG, CL, SHL, AW are with the Department of Computing and Mathematical Sciences, California Institute of Technology, Pasadena,
CA, 91125, USA. Email: \texttt{\{lguo, cliang2, slow, adamw\}@caltech.edu}. AZ is with the Department of Mathematics of the Vrije Universiteit Amsterdam, 1081HV, The Netherlands. Email: \texttt{a.zocca@vu.nl}.}}

\begin{document}

\maketitle

\begin{abstract}
\input{abstract}
\end{abstract}


\section{Introduction}\label{section:intro}
\input{intro}

\section{Preliminaries}\label{section:preliminaries}
\input{preliminaries}


\section{Distribution Factors}
\label{section:DistributionFactors}
\input{DistributionFactors}


\section{Line Failure Localization: Non-cut Outages}\label{section:localization.1}
\input{localization}

\section{Conclusion}\label{section:conclusion}
\input{conclusion}

\bibliographystyle{IEEEtran}
\bibliography{biblio,PowerRef-201202}

\iftoggle{isreport}{
\input{proofs}
}{}

\input{bios}


\end{document}

%% file: abstract.tex
Transmission line failures in power systems propagate non-locally, making the control of the resulting outages extremely difficult. In this work, we establish a mathematical theory that characterizes the 
patterns of line failure propagation and localization in terms of network graph structure.
It provides a novel perspective on distribution factors that precisely captures Kirchhoff's Law in terms of topological structures. Our results show that the distribution of specific collections of subtrees of the transmission network plays a critical role on the patterns of power redistribution, and motivates the block decomposition of the transmission network as a structure to understand long-distance propagation of disturbances.
In Part I of this paper, we present the case when the post-contingency network remains connected after an 
initial set of lines are disconnected simultaneously.
In Part II, we present the case when an outage separates the network into multiple islands.
   

%% file: intro.tex
\begin{figure}[t]
\centering
{
\iftoggle{isarxiv}{
\includegraphics[width=0.4\textwidth]{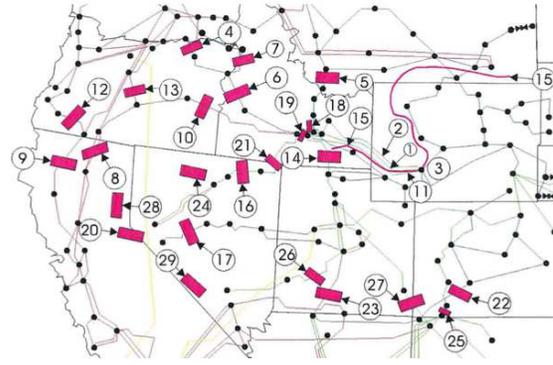}
}
{
\includegraphics[width=0.4\textwidth]{figs/us_blackout.eps}
}
}
\caption{The sequence of events, indexed by the circled numbers, that lead to the Western US blackout in 1996 from \cite{NERC2002, hines2017cascading}.}
\label{fig:blackout}
\end{figure}

Cascading failures in power systems propagate non-locally, making their analysis and mitigation difficult. This fact is illustrated by the sequence of events leading to the 1996 Western US blackout (as summarized in Fig.~\ref{fig:blackout} from \cite{NERC2002, hines2017cascading}), in which successive failures happened hundreds of kilometers away from each other (e.g.~from stage \circled{3} to stage \circled{4} and from stage \circled{7} to stage \circled{8}). Non-local propagation makes it challenging to design distributed controllers that reliably prevent and mitigate cascades in power systems. In fact, such control is often considered impossible, even when centralized coordination is available \cite{bienstock2007integer,hines2007controlling}.

Current industry practice for mitigating cascading failures relies on simulation-based  analysis of creditable contingencies \cite{baldick2008initial}. The size of this contingency set, and thus the level of security guarantee, is often constrained by computational power, undermining its effectiveness in view of the enormous number of components in power networks. After a blackout event, a detailed study typically leads to a redesign of such contingency sets, potentially together with physical network upgrades and revision of system management policies and regulations \cite{hines2007controlling}.

The limitations of the current practice have motivated a large body of research on cascading failures; see e.g. \cite{GuoZheng2017} for a recent review with extensive references. In particular, the literature on analytical properties of cascading failures can be roughly categorized as follows: (a) applying Monte-Carlo methods to analytical models that account for the steady state power redistribution using DC \cite{carreras2002critical,anghel2007stochastic,yan2015cascading,bernstein2014power} or AC \cite{nedic2006criticality,rios2002value,song2016dynamic} power flow models; (b) studying pure topological cascade models built upon simplifying assumptions on the propagation dynamics (e.g., failures propagate to adjacent lines with high probability) and inferring component failure propagation patterns from graph-theoretic properties \cite{brummitt2003cascade,kong2010failure,crucitti2004topological}; (c) investigating simplified or statistical cascading failure dynamics \cite{dobson2005probilistic,wang2012markov,rahnamay2014stochastic,hines2017cascading}.

In all these approaches, it is often difficult to make general inferences about failure patterns. The initial failure could be due to large disturbances in power injections, transmission outages, or human errors~\cite{nesti2018emergent}. The power flow over a transmission line can increase, decrease, and even reverse direction as cascading failure unfolds~\cite{lai2013allerton}. The failure of a line can cause another line that is arbitrarily far away to become overloaded and trip \cite{bernstein2014vulnerability}. The load shedding strategy can increase the congestion on certain lines, instead of mitigating the cascading failure~\cite{bienstock2010n-k}. This lack of structural properties is a key challenge in the modeling, control, and mitigation of cascading failures in power systems.

In this work, we focus on transmission line failures and take a different approach
that leverages the spectral representation of transmission network topology to establish several structural properties. The spectral view is powerful as it reveals surprisingly simple characterizations of complex system behaviors, e.g., on system robustness in terms of effective resistance \cite{ghosh2008minimizing}, on Kron reduction of the power network \cite{dorfler2013kron}, on controllability and observability of power system dynamics \cite{guo2017spectral}, and on monotonicity properties and power flow redistribution \cite{guo2017monotonicity}.

\textbf{Contributions of this paper:} \emph{We establish a mathematical theory that characterizes line failure localization properties of power systems.} This theory makes crucial use of the weighted Laplacian matrix of a transmission network and its spectral properties. Our results unveil a deep connection between power redistribution patterns and the distribution of different families of (sub)trees of the power network topology. 
We show how specific topological structures naturally emerge in the analysis of several important and well-studied quantities in power system contingency analysis, such as the generation shift sensitivity factors and the line outage redistribution factors. 
Further, in contrast to pure graphical models such as those in \cite{motter2002attack,brummitt2003cascade,kong2010failure}, our topological interpretations do not rely on any simplifications on failure propagation but capture Kirchhoff's and Ohm's Law in a precise way under the steady state DC power flow model.

In Part I of this paper, we restrict our attention to the case where the network remains connected after the line failures and study how such failures impact the branch flows on the surviving network. In this scenario the power injections remain balanced and do not need to be adjusted after the contingency. 
Our theory reveals a decoupling structure of the transmission network that leads to failure localizability in a cascading process. More specifically, we show that when a non-cut set of lines trips, any other line that is not in the same block (see Definition \ref{ch:cf; subset:lodf; def:blocks}) of one of the tripped  lines is not affected. In other words, non-cut failures cannot cross the boundaries of the block decomposition of the transmission network. In Part II, we consider the scenario where the set of tripped lines disconnects the network into two or more connected components, called islands, and injections must be adjusted to rebalance the power injections.

The key result that relates the power redistribution to graphical structures is given in Theorem \ref{prop:prem_basis}, which states that the distribution of specific collections of subtrees of the transmission network fully determines the system state under a given set of injections. We then establish a new set of graphical representations of generation shift sensitivity factors and line outage distribution factors in contingency analysis. This novel graph-theoretical viewpoint enables us to derive precise algebraic properties of power redistribution using purely graphical argument, and shows that disturbances propagate through ``subtrees'' in a power network. Using this framework, in Section \ref{section:localization.1} we derive the \textit{Simple Cycle Criterion} that precisely determines whether the failure of one line can impact another line in a given network and fully characterizes non-cut failure propagation. 

In order to prove these results, we exploit the celebrated Kirchhoff Matrix Tree Theorem as well as its generalization to all matrix minors. Further, we make use of some novel properties of the Laplacian matrix derived in the context of DC power flow and express various quantities of interest, including distribution factors, using specific collection of (sub)trees of the the transmission network.

The results here can be extended in several directions.  For example, injection disturbances such as loss of generators or loads can be readily incorporated into the same framework as initial failures.
In \cite{LiangGuo2020,zocca2020spectral}, we explore ways to judiciously 
switch off a small number of transmission lines to create more blocks to enhance failure localization
and develop real-time mitigation strategy.
This technique can be  synergistically applied, or sometimes replace, controlled islanding
(see e.g. 
 \cite{Vittal1998, Vittal2003, XuVittal2010, Sun2003, Li2005, Li2010, Ding2013, Bialek2014, Kirschen2019})
 as a corrective action, in which an inter-connected power system will be partitioned into multiple blocks after a contingency that are connected by either bridges or cut vertices\footnote{We thank Janusz Bialek for suggesting this as a potential application.}. By not separating the system into multiple islands, more loads can potentially be supported in the emergency state, more reliably, until restoration.

%% file: preliminaries.tex
In this section, we introduce our model for line failure cascading based on standard DC power flow equations. 


\subsection{DC Power Flow Cascading Model}
We describe a transmission network using a graph $\calG=(\calN,\calE)$, whose node set $\calN=\set{1,\ldots, n}$ models the $n=|\calN|$ buses and whose edge set $\calE\subseteq \calN \times \calN$ models the $m=|\calE|$ transmission lines. We use the terms bus/node and line/edge interchangeably. An edge in $\calE$ between node $i$ and $j$ is denoted either as $l$ or $(i,j)$. Without loss of generality, we assume the graph is simple and we assign an arbitrary orientation to the edges in $\calE$ so that if $(i,j)\in\calE$ then $(j,i)\notin\calE$. The susceptance (weighted by nodal voltage magnitudes) of edge $l=(i,j)$ is denoted as $B_l=B_{ij}$ and the susceptance matrix is the $m\times m$ diagonal matrix $B:=\diag(B_l: l\in\calE)$. The incidence matrix of $\calG$ is the $n\times m$ matrix $C$ defined as
$$
C_{il}=\begin{cases}
  1 & \text{if node }i\text{ is the source of }l,\\
  -1 & \text{if node }i\text{ is the target of }l,\\
  0 &\text{otherwise.}
\end{cases}
$$
Let $f$ be the $m$-dimensional vector consisting of all branch flows, with $f_l$ denoting the flow on edge $l$. We introduce the $n$-dimensional vectors $p$ and $\theta$, where $p_i$ and $\theta_i$ are the power injection and voltage phase angle at node $i$, respectively. With the above notation, the DC power flow model is described by the following equations
\begin{subequations}\label{eqn:dc_model}
\begin{IEEEeqnarray}{rCl}
	p&=&Cf, \label{eqn:flow_conservation}\\
	f&=&BC^T\theta, \label{eqn:kirchhoff}
\end{IEEEeqnarray}
\end{subequations}
where \eqref{eqn:flow_conservation} is the flow conservation (Kirchhoff's) law and \eqref{eqn:kirchhoff} is the Ohm's laws. 
Given an injection vector $p$ that is \textit{balanced} over the network, i.e., $\sum_{j\in\calN}p_j=0$, the DC model \eqref{eqn:dc_model} has a solution $(\theta, f)$ that is unique up to an arbitrary reference angle.  Without loss of generality, we choose node $n$ as a reference node and set $\theta_n=0$. Using this convention, the solution $(\theta, f)$ is unique.

We consider a cascading failure process that starts with an initial set of line outages 
that may sequentially cause more line outages.
At any stage of the cascade, if a set $F \subset \calE$ of transmission lines is tripped, then  power redistributes according to the DC model \eqref{eqn:dc_model} on the post-contingency graph $\calG':=(\calN,\calE\bs F)$. We assume that each transmission line $l$ has a steady-state thermal capacity $\pi_l$ so that if the current branch flow $|f_l|$ exceeds $\pi_l$, the line is tripped in the next stage, causing another power flow redistribution and possibly further failures. The cascade stops when all branch flows are below their capacities. 


\subsection{Laplacian Matrices and Power Flow Equations}
The DC power flow equations \eqref{eqn:dc_model} imply that
$$
p  = CBC^T\theta  =:  L \theta,
$$
where $L=CBC^T\in \R^{n\times n}$ is called the Laplacian matrix of $\calG$~\cite{chung1997spectral}.
It is well-known that $L$ is a symmetric and positive semi-definite matrix with zero row sums.
If $\calG$ is connected, then $L$ is of rank $n-1$, its null space is spanned by the vector $\textbf{1} =(1,\dots,1) \in \R^{n}$, and the Penrose-Moore pseudo-inverse of $L$ is the $n\times n$ matrix
\begin{equation*}
L^\dag  :=  \left( L + \textbf 1 \textbf 1^T/n \right)^{-1} - \textbf 1 \textbf 1^T/n.
\end{equation*}
Given an injection vector $p$ that is balanced, i.e., $\sum_i p_i = 0$, a power flow 
solution can be written in terms of $L^{\dag}$  as $\theta = L^\dag p$ and $f = BC^TL^\dag p$. 
This formulation yields unique branch flows $f$ and phase angles $\theta$. However, it may not satisfy the aforementioned convention prescribing the reference phase angle to be zero, as it may be that $\theta_n \neq 0$. Let the reduced Laplacian matrix $\ol{L}$ be the submatrix of $L$ obtained by deleting its  $n$-th row and column (corresponding to the reference node). If the network is connected, then $\ol L$ is invertible and we can define an $n\times n$ matrix $A$ by 
\begin{equation}\label{eqn:extension_of_a}
A= \begin{bmatrix} \paren{\ol{L}}^{-1} & \bm{0} \\ \bm{0} & 0 \end{bmatrix}.
\end{equation}
Given a balanced injection vector $p$, the power flow solution can also be written in terms of $A$  as $\theta' = Ap$ and $f' = BC^TAp$. In this representation, the reference phase angle satisfies $\theta_n'=0$. Moreover, we have $\theta' = \theta -  \theta_n \textbf 1$, i.e., the two phase angle vectors differ by a constant reference angle. It should be noted that the branch flow vector is always unique, $f=f'$.

%
\subsection{Block Decomposition}

The failure localizability of a network depends critically on the notion of blocks in graph theory~\cite{harary1969graph}.
Every graph can be uniquely and efficiently decomposed into blocks, which are its maximal 2-connected components. 
%
Formally, let $\mathfrak R$ be the relation on the edges of $\calG$ defined by $l_1 \mathfrak R l_2$ if (and only if) $l_1 = l_2$ or they belong to a common simple cycle of $\calG$\footnote{A cycle is simple if the only repeated vertex is the first/last one.}. It can be shown that $\mathfrak R$ is an equivalence relation on the set of edges $\calE$, inducing network blocks as follows.

\begin{figure}[t]
\centering
\subfloat[]{\includegraphics[width=.22\textwidth]{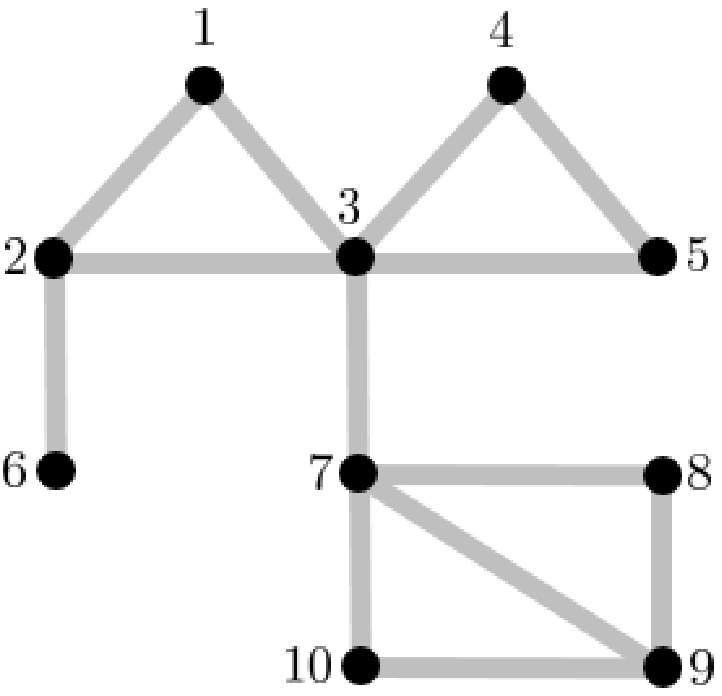}
}\hfill
\subfloat[]{\includegraphics[width=.22\textwidth]{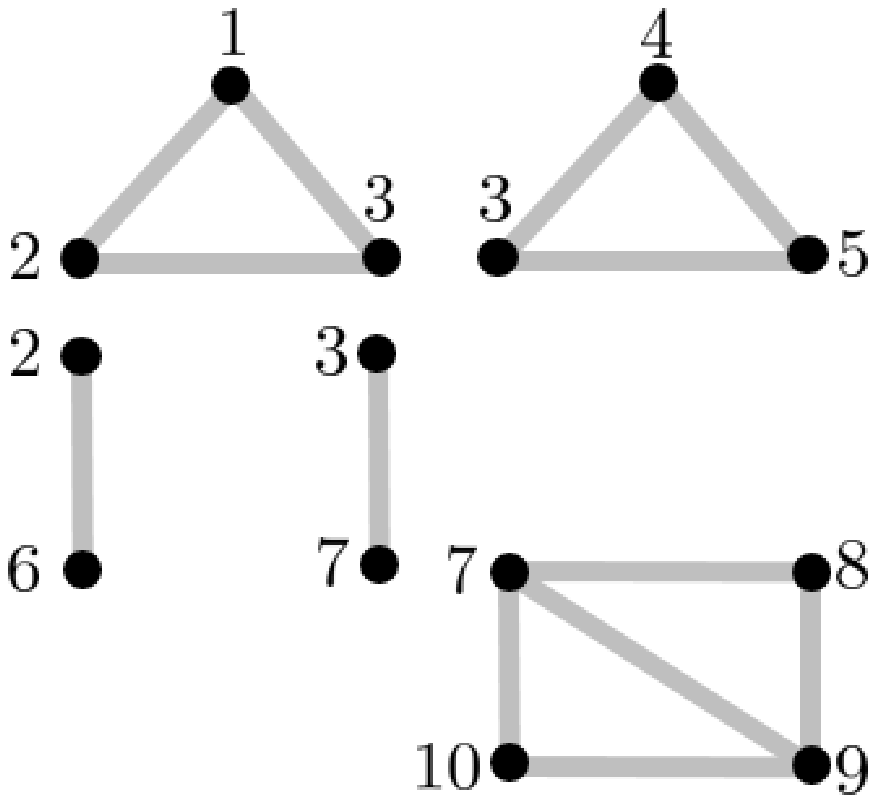}
}
\caption{(a) An undirected graph $\calG$.  (b) The block decomposition of $\calG$. Vertices $2, 3, 7$ are cut vertices of $\calG$.  Edges $(2, 6)$ and $(3, 7)$ are bridges.}
	\label{ch:cf; sec:treepartitioning; fig:blocks}
\end{figure}

\begin{defn}[Blocks, bridges, cut vertices]$ $
\label{ch:cf; subset:lodf; def:blocks}
\begin{enumerate}
\item The subgraphs of $\calG$ induced by the equivalence classes of $\mathfrak R$ are called \emph{\textbf{blocks}} of $\calG$.
\item A node of $\calG$ that is part of two or more blocks is called a \emph{\textbf{cut vertex}} of $\calG$.
\item An edge in a singleton equivalent class is called a \emph{\textbf{bridge}} of $\calG$. A block that is not a bridge is called a \emph{\textbf{non-bridge}} block.
\item A subset $F\subseteq \calE$ of edges is called a \emph{\textbf{cut set}} of $\calG$ if removing all edges in $F$ disconnects the graph. A set $F\subseteq \calE$ is called a \emph{\textbf{non-cut set}} if $F$ is not	a cut set.
\end{enumerate}
\end{defn}
The removal of a cut vertex disconnects $\calG$.  A bridge is a cut set of size one, since its removal disconnects $\calG$. Two non-bridge blocks are connected either by a bridge or by a cut vertex. These definitions are illustrated in Figure \ref{ch:cf; sec:treepartitioning; fig:blocks}.

The block decomposition of a graph $\calG$ is unique and there exist efficient algorithms to find all blocks of a graph $\calG$ that run in $O(|\calN|+|\calE|)$ in time and space on a single processor or run in $O(\log |\calN|)$ in time and $O(|\calN|+|\calE|)$ in space on $O(|\calN|+|\calE|)$ processors~\cite{Tarjan1985}.

%% file: DistributionFactors.tex
In this section we focus on  \textit{distribution factors} widely used in contingency analysis, and derive novel expressions for them in terms of network graph structures. We also explain the implication of this spectral representation of network graph on distribution factors. 
In Section \ref{section:localization.1} and Part II of this paper, we use these results to reveal the decoupling structure of distribution factors and the resulting failure localization property of a power network. 

\subsection{Graphical Interpretation}
We first introduce some additional notations useful to work with  \textit{spanning trees} of the graph $\calG$ and present a preliminary result (Theorem \ref{prop:prem_basis}) that gives an interpretation of matrix $A$ in terms of tree structures in $\calG$. 

\begin{figure}[!h]
\centering
\iftoggle{isarxiv}{
\includegraphics[width=.3\textwidth]{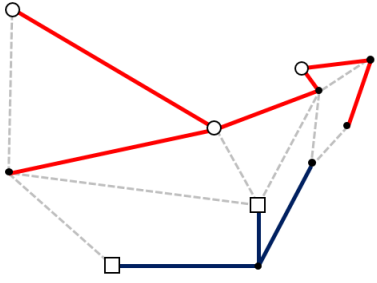}
}{
\includegraphics[width=.3\textwidth]{figs/spanning_forest.eps}
}
\caption{An example element in $\calT(\calN_1,\calN_2)$, where circles correspond to elements in $\calN_1$ and squares correspond to elements in $\calN_2$. The two trees containing $\calN_1$ and $\calN_2$ are highlighted as solid lines.}\label{fig:spanning_forest}
\end{figure}

Given a subset $F \subseteq \calE$ of edges, we denote by $\calT_F$ the set of spanning trees of $\calG$ with edges from $F$ and by $\calT_{-F}$ the set of spanning trees with edges from $-F:=\calE \setminus F$. 
In particular, $\calT_\calE$ is the set of all spanning trees on $\calG$. 
For any pair of subsets $\calN_1, \calN_2\subset\calN$, we define $\calT(\calN_1, \calN_2)$ to be the set of spanning forests of $\calG$ consisting of exactly two disjoint trees that contain $\calN_1$ and $\calN_2$, respectively (see Fig.~\ref{fig:spanning_forest}). By definition, $\calT(\calN_1,\calN_2)=\emptyset$ if $\calN_1 \cap \calN_2 \neq \emptyset$.
To further simplify notations, we omit the braces when there is no confusion, e.g. we write $\calT(ij, \hat i \hat j)$ for $\calT(\set{i,j}, \{\hat i, \hat j \})$ and $\calT_{-l}$ for $\calT_{-\set{l}}$. Given a subset $F \subseteq \calE$ of edges, we define its weight as
$$
\beta(F):=\prod_{l\in F}B_l.
$$
Note that $\beta(F) >0$ since, by construction, the susceptances $B_l$, $l \in \calE$, are all positive.
%
%

The Kirchhoff's Matrix Tree Theorem relates the determinant of the reduced Laplacian matrix $\ol L$ and its minors to the total weight of (a specific collection of) the spanning trees of $\calG$ \cite{chaiken1982allminor}.
\FloatBarrier
\begin{lemma}[Matrix Tree Theorem \cite{chaiken1982allminor}]$ $
\label{prop:prem_laplace_tree}
\begin{enumerate}
\item The determinant of $\ol{L}$ is given by
$$
\det(\ol{L})=\sum_{F\in\calT_\calE} \beta(F).
$$
\item The determinant of the matrix $\ol{L}^{ij}$ obtained from $\ol{L}$ by deleting the $i$-th row and $j$-th column is given by
  $$
  \det\paren{\ol{L}^{ij}}=(-1)^{i+j}\sum_{F\in \calT(ij, n)}\beta(F).
  $$
\end{enumerate}
\end{lemma}


Lemma \ref{prop:prem_laplace_tree} leads to our first main result, proved in Appendix~\ref{section:spectral_rep_proof}, that provides a graphical interpretation of the entries of the matrix $A$. It underlies the decoupling structure of distribution factors and failure localizability of network graph presented in the rest of this two-part paper.

\begin{thm}[Spectral Representation]\label{prop:prem_basis}
If $\calG$ is connected, then for any pair of nodes $i,j\in\calN$, 
we have
  \begin{equation}\label{eqn:atom_rep}
  A_{ij}=\frac{\sum_{F\in \calT(ij, n)}\beta(F)}{\sum_{F\in \calT_\calE}\beta(F)}.
  \end{equation}
\end{thm}
The denominator in~\eqref{eqn:atom_rep} is a normalization constant common for all entries of $A$. The sum in the numerator is over all  trees in $\calT(ij, n)$, which means that $A_{ij}$ is proportional to the (weighted) number of trees that connect $i$ to $j$ without traversing the reference node $n$, and can be interpreted as the ``connection strength'' between the nodes $i$ and $j$ in $\calG$.

Since $A$ determines the power flow solution $(\theta, f)$, Theorem~\ref{prop:prem_basis} allows us to deduce analytical properties of a DC solution using its graph structure. In particular, it provides new graph theoretic expressions for distribution factors.

\subsection{Power Transfer Distribution Factor (PTDF)} \label{section:gdf}
\begin{figure}[t]
\centering
\iftoggle{isarxiv}{
\includegraphics[width=.3\textwidth]{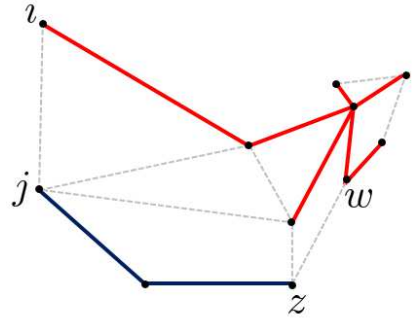}
}{
\includegraphics[width=.3\textwidth]{figs/spectral_correlation.eps}
}
\caption{An example element in $\calT(i\hat{i},j\hat{j})$. The spanning trees containing $\{i,\hat{i}\}$ and $\{j,\hat{j}\}$ are highlighted as solid lines. }\label{fig:spectral_correlation}
\end{figure}

Consider a pair of buses $\hat i$ and $\hat j$, not necessarily adjacent in the graph $\calG$. Suppose the injection at bus $\hat i$ is increased by $\Delta_{\hat i\hat j}$, the injection at bus $\hat j$ is reduced 
by $\Delta_{\hat i \hat j}$, and all other injections remain unchanged so that the new injections remain balanced.
Let $f_{l}$ and $\tilde f_{l}$ denote the branch flow on any line $l\in \calE$ before and after the injection change (both uniquely determiend by the DC power flow equations \eqref{eqn:dc_model}) and let $\Delta f_{l} := \tilde f_{l} - f_{l}$ be their difference.
The \textit{power transfer distribution factor (PTDF)}, also known as \textit{generation shift sensitivity factor}, is defined as~\cite{wood1996generation}:
$$
D_{l,\hat i\hat j} := \frac{\tilde f_{l} - f_{l}} {\Delta_{\hat i \hat j}} = \frac{\Delta f_{l}}{\Delta_{\hat i\hat j}}.
$$
The factor $D_{l, \hat i \hat j}$ can be explicitly computed in terms of  matrix $A$ (letting $l=(i, j)$) \cite{wood1996generation}:
$$
D_{l,\hat i \hat j} = B_{l} (A_{i\hat i} + A_{j\hat j} - A_{i\hat j}- A_{j\hat i}).
$$
Applying Theorem~\ref{prop:prem_basis} to this formula yields the following result (\iftoggle{isreport}{proved in Appendix \ref{section:prem_proof_of_cross}}{see our online report \cite{part1} for a proof}).
\begin{thm}\label{cor:prem_cross_product}
If $\calG$ is connected, then for any pair of nodes $\hat i,\hat j\in\calN$ and any edge $l=(i,j)\in\calE$, we have
  \begin{align*}
D_{l, \hat i \hat j} = \frac{B_l
  \left(\sum_{F\in \calT(i\hat i, j \hat j)}\beta(F)
  -\sum_{F\in \calT(i \hat j, j \hat i)}\beta(F) \right)}{\sum_{F\in \calT_\calE}\beta(F)} .
  \end{align*}
\end{thm}
Despite its apparent complexity, this formula carries an intuitive graphical meaning. The two sums in the numerator are over the spanning forests $\calT(i\hat i, j\hat j)$ and $\calT(i \hat j, j \hat i)$. Each element in $\calT(i\hat i, j \hat j)$, as illustrated in Fig.~\ref{fig:spectral_correlation}, specifies a way to connect $\hat i$ to $i$ and $\hat j$ to $j$ through disjoint trees and represents a possible path for buses $\hat i,\hat j$ to ``spread'' impact to line $(i, j)$. Similarly, elements in $\calT(i\hat j,j\hat i)$ represents possible paths for buses $\hat i,\hat j$ to ``spread'' impact to $(j,i)$, which counting orientation, contributes negatively to line $l$. Theorem \ref{cor:prem_cross_product} thus implies that the impact of shifting generations from $\hat j$ to $\hat i$ propagates to the line $l=(i,j)$ through all possible spanning forests that connect the endpoints $\hat i, \hat j, i, j$ (accounting for orientation). The relative strength of the trees in these two families determines the sign of $D_{l, \hat i\hat j}$.

If $\hat l=(\hat i,\hat j)\in\calE$ is also a transmission line in the grid, we use the more compact notation $D_{l\hat l}$ for $D_{l,\hat i\hat j}$ and introduce the $m\times m$ PTDF matrix $D := (D_{l \hat l}, l, \hat l \in \calE)$.
Corollary \ref{ch:lm; corollary:PTDF.2} summarizes how the PTDF matrix $D$ can be expressed explicitly in terms of matrix $A$.
\begin{cor}
\label{ch:lm; corollary:PTDF.2}
Assume $\calG$ is connected. Then,
\begin{enumerate}
\item $D = B C^T A C$.

\item For each line $l\in \calE$ the corresponding diagonal entry of $D$ is given by:
	\begin{eqnarray*}
	D_{ll} = 1 - \frac{ \sum_{F\in \calT_{-l}} \beta(F) } { \sum_{F\in \calT_\calE} \beta(F) }.
	\end{eqnarray*}
	Hence, $D_{ll} = 1$ if $l$ is a bridge and $0<D_{ll}<1$ otherwise.
\end{enumerate}
\end{cor}
 This corollary is a direct consequence of Theorem \ref{cor:prem_cross_product} and we omit its proof.

\subsection{Line Outage Distribution Factor (LODF)}\label{section:lodf}
The line outage distribution factor (LODF) 
describes the impact of line outages on the power flows in the post-contingency network.
We call the contingency a \emph{non-cut ($F$) outage} if a non-cut set $F$ of lines trip simultaneously, and
a  \emph{non-bridge ($\hat l$) outage} 
when the non-cut set $F = \{\hat l\}$ is a singleton.
We first study the impact of a non-bridge outage and then generalize it to a non-cut set outage.

%
\subsubsection{Non-bridge $\hat l$ outage}

The line outage distribution factor (LODF) $K_{l \hat l}$ is defined to be the branch flow change $\Delta f_{l}$ on a post-contingency surviving line $l$ when a \emph{single non-bridge} line $\hat l$ trips, normalized by the pre-contingency branch flow $f_{\hat l}$ over the tripped line:
\begin{eqnarray*}
K_{l \hat l} := \frac{\Delta f_{l}}{f_{\hat l}}, \quad l \neq \hat l \in \calE, \ \text{ non-bridge line } \hat l \in \calE,
\end{eqnarray*}
assuming that the injections  remain unchanged since the network remains connected.
Writing $l=(i,j), \hat{l}=(\hat i,\hat j)$, $K_{l \hat l}$ can be calculated as \cite{wood1996generation}:
\begin{equation}
K_{l \hat l} \ = \ 
\frac{B_{l} \left( A_{i\hat i}+A_{j \hat j}-A_{i \hat j} - A_{j \hat i} \right) }{ 1 - B_{\hat l} (A_{\hat i\hat i}+A_{\hat j\hat j}-A_{\hat i\hat j}-A_{\hat j\hat i})}
\ = \ \frac{D_{l \hat l}} {1 - D_{\hat l \hat l} },
\label{partI; eq:KandD}
\end{equation}
which is independent of the power injections. This formula only holds if the post-contingency graph $\calG':=(\calN,\calE\bs\{\hat l\})$ is connected, as otherwise its denominator is $0$ by Corollary \ref{ch:lm; corollary:PTDF.2}.
Combining Theorem \ref{cor:prem_cross_product} and Corollary \ref{ch:lm; corollary:PTDF.2} immediately yields the following new formula for $K_{l \hat l}$.

\begin{thm}\label{thm:redist_factor}
Let $\hat{l}=(\hat i,\hat j)$ be an edge such that $\calG':=(\calN,\calE\bs\{\hat l\})$ is connected. Then, for any other edge $l=(i,j)$,
\begin{equation}
K_{l \hat l} = \frac{B_{l}
\left( \sum_{F\in \calT(i\hat i, j\hat j)}\beta(F) - \sum_{F\in \calT(i \hat j, j \hat i)}\beta(F) \right)} {\sum_{F\in\calT_{-\hat l}}\beta(F)} . 
\label{eqn:redist_factor}
\end{equation}
\end{thm}
As in Theorem~\ref{cor:prem_cross_product}, each term of \eqref{eqn:redist_factor} also carries clear graphical meanings: (a) The numerator of \eqref{eqn:redist_factor} quantifies the impact of tripping $\hat l$ propagating to $l$ through all possible trees that connect $\hat l$ to $l$, counting orientation. (b) The denominator of \eqref{eqn:redist_factor} sums over all spanning trees of $\calG$ that do not pass through $\hat l=(\hat i,\hat j)$, and each tree of this type specifies an alternative path that power can flow through if line $\hat l$ is tripped. When there are more trees of this type, the network has a better ability to ``absorb'' the impact of line $\hat l$ being tripped, and the denominator of \eqref{eqn:redist_factor} precisely captures this effect by saying that the impact of $\hat l$ being tripped on other lines is inversely proportional to the sum of all alternative tree paths in the network. (c) The susceptance $B_{l}$  in \eqref{eqn:redist_factor} captures the intuition that lines with smaller susceptance tend to be less sensitive to power flow changes from other parts of a power network.

\subsubsection{Non-cut $F$ outage}
We now extend the results for LODFs from a non-bridge outage to a non-cut outage.
Let $F\subsetneq \calE$ be a non-cut set of lines that are disconnected simultaneously and $m_F:=|F|$ be the number of disconnected lines. Denote by $-F := \calE \! \setminus \! F$ the set of surviving lines and assume that the injections $p$ remains unchanged.
Partition the susceptance matrix $B$ and the incidence matrix $C$ into submatrices corresponding to surviving lines in $-F$ and tripped lines in $F$:
\begin{eqnarray}
B =: \begin{bmatrix} B_{-F} & 0 \\ 0 & B_{F} \end{bmatrix}, \quad C =: [C_{-F} \ \ C_{F}].
\label{ch:cf; subsec:LODF; eq:BCpartitions.1}
\end{eqnarray}
Similarly, we can partition the PTDF matrix $D=(D_{l\hat l}, l, \hat l \in \calE)$ into submatrices corresponding to non-outaged lines in $-F$ and outaged lines in $F$, possibly after permutations of rows and columns\footnote{We also write $D_{F, \hat F}$ as $D_{F\hat F}$ when there is no confusion.}:
\begin{eqnarray*}
D = \begin{bmatrix} D_{-F, -F} & D_{-FF} \\ D_{F, -F} & D_{FF} \end{bmatrix}.
\end{eqnarray*}
Since $D = B C^T A C$ from Corollary \ref{ch:lm; corollary:PTDF.2},
we have
\begin{eqnarray*}
D =\begin{bmatrix} B_{-F}C_{-F}^T A C_{-F} & B_{-F}C_{-F}^T A C_{F} \\ B_{F}C^T_{F}AC_{-F} & B_{F}C^T_{F}AC_{F}\end{bmatrix}.
\end{eqnarray*}
Similarly to the case of non-bridge outage, the post-contingency flow changes $\Delta f_{-F}:=(\Delta f_l, l\in -F)$ on the surviving lines depend linearly on the pre-contingency branch flows $f_F:=(f_{\hat l}, \hat l\in F)$ on the tripped lines.  The sensitivities of $\Delta f_{-F}$ to $f_F$ implicitly define a $(m-m_F)\times m_F$ matrix $K^F := K^F_{-FF} := (K^F_{l \hat l}, l \in -F, \hat l \in F)$, called the \textit{Generalized LODF (GLODF)} with respect to a non-cut $F$ outage, namely
\begin{eqnarray}
\Delta f_{-F} = K^F \, f_F,	\quad \text{for non-cut set }F \subsetneq \calE.
\label{ch:cf; subsec:lodf; eq:defKlF}
\end{eqnarray}
If we stack the LODF for single line outages into matrix $K_{-FF}:=(K_{l\hat l}, l\in -F, \hat l \in F)$, $K_{-FF}$ has the same dimension as the GLODF $K^F$ for a non-cut $F$ outage. Note that every element $K_{l\hat l}$ is the LODF when single non-bridge lines are tripped, as derived in~\eqref{partI; eq:KandD}. Mathematically, we can write $K_{-FF}$ in the matrix form:
\begin{eqnarray}
K_{-FF} = D_{-FF} \left( I - \text{diag}(D_{FF}) \right)^{-1}, \label{eqn:K_-FF}
\end{eqnarray}
where $\text{diag}(D_{FF}) := \text{diag}(D_{ll}, l\in F)$. In general $K^F \neq K_{-FF}$. However, they are related in the next result.
Let $L_{-F}:=C_{-F}B_{-F}C_{-F}^T$ be the Laplacian matrix of the post-contingency network.
Let $\overline{L}_{-F}$ be the submatrix of $L_{-F}$ obtained by deleting its $n$-th row and column, and $A_{-F} :=\begin{bmatrix} (\overline{L}_{-F})^{-1} & \bm{0} \\ \bm{0} & 0 \end{bmatrix}$.

\begin{thm} [GLODF $K^F$ for non-cut outage]
\label{ch:cf; thm:LODF.3}
Suppose a non-cut set $F\subsetneq \calE$ of lines trip simultaneously so that the surviving graph $(\calN, \calE\!\setminus\! F)$ remains connected.   
\begin{enumerate}
\item The GLODF $K^F$ defined in \eqref{ch:cf; subsec:lodf; eq:defKlF} is given in terms of post-contingency network by:
\begin{subequations}
	\begin{eqnarray}
	K^F = B_{-F} C_{-F}^T\, A_{-F}\, C_F,
	\label{ch:cf; eq:K^F.2}
	\end{eqnarray}

\item $K^F$ is given in terms of pre-contingency inverses by:
\begin{eqnarray}
K^F  = D_{-FF} \left( I - D_{FF} \right)^{-1},
\label{ch:cf; eq:GLODF.1a}
\end{eqnarray}
or, equivalently,
\begin{eqnarray}
K^F	 =  B_{-F} C_{-F}^T  A  C_F
	\left( I - B_{F} C_{F}^T A C_F \right)^{-1}.
\label{ch:cf; eq:GLODF.1b}
\end{eqnarray}
The matrix
$I - D_{FF} \ = \ I - B_{F}  C_{F}^T  A  C_F$ 
is invertible provided $F$ is a non-cut set of disconnected lines.
\item $K^F$ is related to the LODF matrix $K_{-FF}$ when single non-bridge lines are outaged through:
\begin{eqnarray}
K^F = K_{-FF} \left( I - \diag(D_{FF}) \right) \left( I - D_{FF} \right)^{-1}.
 \label{ch:cf; eq:GLODF.1c}
 \end{eqnarray}
\label{ch:cf; eq:GLODF.1}
\end{subequations}
\end{enumerate}
\end{thm}
%
%
Theorem \ref{ch:cf; thm:LODF.3} is proved in Appendix \ref{appendix:proof_of_glodf}. The formulae \eqref{ch:cf; eq:GLODF.1a}-\eqref{ch:cf; eq:GLODF.1b} generalize \eqref{partI; eq:KandD} from a non-bridge outage to a non-cut set outage.
As mentioned earlier, $K^F \neq K_{-FF}$ unless $F = \{\hat l\}$ is a singleton and \eqref{ch:cf; eq:GLODF.1c} clarifies their relationship. This fact shows that the impact for multiple simultaneous line outages is not a simple superposition of the corresponding single line outages, as their effects are coupled  by power flow physics and network topology.

\subsection{Remarks}
The reference \cite{AlsacStottTinney1983} seems to be the first to introduce the use of matrix inversion lemma to power systems to study the impact of network changes on line currents resulting from Ohm's law $I = YV$ where $Y$ is a network Laplacian matrix, e.g., nodal admittance or Jacobian matrix from the linearization of AC power flow equations.\footnote{LODF for multi-line outages is also developed in, e.g., \cite{EnnsQuadaSackett1982}, but without the simplification of the matrix inversion lemma.}  
The changes can be changes in the line parameters $Y_{ij}$ or outages of an arbitrary set of lines, or changes in the nodal injections $I$ or outages of an arbitrary set of generators.
This linear system is mathematically equivalent to the DC power flow model \eqref{eqn:dc_model}. In \cite{StottAlsacAlvarado1985}, the method of \cite{AlsacStottTinney1983} is applied to the DC power flow model to characterize the flow change for an arbitrary set of line outages.
The paper  \cite{StottAlsacAlvarado1985} also allows generator outages and use these formulae to quickly rank contingencies in security analysis. 
The formula \eqref{ch:cf; eq:GLODF.1b}, called generalized LODF (GLODF), is re-discovered in  \cite{GulerGrossLiu2007, GulerGross2007} using a different method, likely
 unaware of the results of \cite{AlsacStottTinney1983, StottAlsacAlvarado1985}. The underlying idea of the letter \cite{GulerGrossLiu2007} is to emulate line outages through changes in injections on the pre-contingency network by judiciously choosing injection at the tail of each disconnected line and withdrawal at its head using PTDF, starting from the expression \eqref{partI; eq:KandD} for single-line outage and proving the general non-cut set case by induction. The paper  \cite{GulerGross2007} uses the relation between $\Delta f_{-F}$ and PTDF to detect islanding: $F$ is a cut set that disconnects the network if and only if the inverse in \eqref{ch:cf; eq:GLODF.1a} ceases to exist. See also \cite{GuoShahidehpour2009} for another derivation of GLODF in terms of PTDF. While PTDF and LODF determine the sensitivity of power flow solutions to parameter changes, one can also study the sensitivity of optimal power flow solutions to parameter changes; see, e.g., \cite{Gribik1990, Hauswirth2018}. LODFs are also studied more recently as a tool to quantify network robustness and flow rerouting \cite{Witthaut2019}.

%% file: localization.tex
In this section, we first introduce the Simple Cycle Criterion that characterizes whether the branch flow on a surviving line is impacted by a non-bridge outage. 
We then use it to explain failure localizability of a power network: for a non-cut set outage, the impact is localized within each block where outages occur.

\subsection{Simple Cycle Criterion}
\label{section:SimpleLoopC}

Theorem \ref{thm:redist_factor} shows that whether the tripping of a line $\hat l$ will impact another line $l$ or not depends on how these two lines are connected by subtrees of $\calG$.  
We now establish a simple criterion that can be directly verified on the network graph. It states that the outage of line $\hat l$ will impact the branch flow on line $l$, i.e., $K_{l\hat l} \neq 0$, only if there is a simple cycle in $\calG$ that contains both lines (recall that a cycle in a graph $\calG$ is called a \emph{simple cycle} if it visits each vertex at most once except for the first/last vertex).

The converse holds ``almost surely'' in the following sense. Suppose the line susceptances are specified by a random vector $B + \omega := (B_l + \omega_l, l\in \calE)$ where the random vector $\omega := (\omega_l, l\in \calE)$ is drawn from a multidimensional probability measure $\mu$ that is absolutely continuous with respect to the Lebesgue measure ${\cal L}_m$, i.e., for any measureable set $X$, ${\cal L}_m(X) = 0$ implies $\mu(X)=0$. By the Radon-Nikodym Theorem~\cite{rudin1987real}, the probability measure $\mu$ is absolutely continuous with respect to $\calL_m$ if and only if it has a probability density function. This essentially amounts to requiring the measure $\mu$ to not contain Dirac masses. In practice, such random vector $\omega$ can model manufacturing, measurement, or modeling errors.
For two predicates $s_1 := s_1(B+\omega)$ and $s_2$ with $s_1$ dependent on the value of the random vector $B+\omega$, we say \emph{$s_1$ ``if'' and only if $s_2$} when $s_1$ implies $s_2$ and $s_2$ almost surely implies $s_1$, or mathematically, we have
\begin{equation*}
s_1\ \Rightarrow \ s_2 \quad \text{ and } \quad 
s_2 \ \Rightarrow \ \mu(s_1(B+\omega)) = 1.
\end{equation*}

\begin{thm}[Simple Cycle Criterion]\label{prop:simple_loop}
For any $\hat{l}=(\hat i,\hat j)\in\calE$ such that $\calG':=(\calN,\calE\bs\{\hat l\})$ is connected and $l=(i,j) \in \calE$, we have $K_{l \hat l}\neq 0$ ``if'' and only if there exists a simple cycle in $\calG$ that contains both $l$ and $\hat{l}$.
\end{thm}

We refer the readers to \cite{guo2018cdc} for an explicit zero probability example where the \emph{``if''} part does not follow.

\subsection{Localization of Non-cut Outages}
We now use Theorem \ref{prop:simple_loop} (proved in Appendix~\ref{proof:SimpleLoopC}) to explain failure localizability of the network graph $\calG$ using its unique block decomposition (see Definition \ref{ch:cf; subset:lodf; def:blocks}).
Recall that two distinct edges are in the same block if and only if there is a simple cycle that contains both of them. Theorem \ref{prop:simple_loop} then implies the following failure localization property when a single non-bridge line $\hat l$ trips: only surviving lines in the same block as $\hat l$ may see their branch flows impacted. In particular, since a bridge is a block, a non-bridge outage will not impact the branch flow on any other bridge. Additionally, the PTDF matrix $D$ has the same decoupling structure with the block decomposition as $D_{l \hat l} = K_{l \hat l} (1 - D_{\hat l \hat l} )$ from \eqref{partI; eq:KandD}.
From these considerations, the following result readily follows.
\begin{cor}[Failure localization: non-bridge outage] 
\label{ch:cf; corollary:blockpartition}
Suppose a single non-bridge line $\hat l=(\hat i,\hat j)$ trips so that the surviving graph $(\calN, \calE\!\setminus\! \{\hat l\})$ remains connected. For any surviving line $l = ( i, j)$ the following statements hold:
\begin{enumerate}
\item LODF $K_{l \hat l} = 0$ if $l$ and $\hat l$ are in different blocks of $\calG$.
\item PTDF $D_{l \hat l} = 0$  if and only if $K_{l \hat l} = 0$.
\end{enumerate}
\end{cor}

To extend failure localizability to the case of a non-cut $F$ outage we use \eqref{ch:cf; eq:GLODF.1c} in Theorem \ref{ch:cf; thm:LODF.3} to express the GLODF $K^F$ in terms of the LODF and PTDF submatrices $K_{-FF}$ and $D_{FF}$.  
Corollary~\ref{ch:cf; corollary:blockpartition} implies a block-diagonal structure of $K_{-FF}$ and $D_{FF}$ which then translates into the same block-diagonal structure of the GLODF $K^F$. 
Specifically, assume the set $\calE$ of lines consists of $b$ blocks $\calE_k$ such that $\calE = \calE_1 \cup \cdots \cup \calE_b$ and $\calE_j\cap \calE_k = \emptyset$ for $j\neq k$.  
Partition the set $F$ of simultaneously outaged lines into $b$ disjoint subsets 
$F_k := F\cap \calE_k$, $k=1, \dots, b$, such that $F = \cup_k F_k$.   
Similarly partition the set $-F$ of surviving lines into $b$ disjoint subsets
$F_{-k} := -F\cap \calE_k = \calE_k \!\setminus\! F_k$, $k=1, \dots, b$, such that $-F = \cup_k F_{-k}$.   
Without loss of generality we assume that the lines are indexed so that the outaged lines in $F_1$ correspond to the first
$|F_1|$ columns of $K_{-FF}$, the outaged lines in $F_2$ correspond to the following $|F_2|$ columns of $K_{-FF}$, so on and so forth, and the tripped lines in $F_b$ correspond to the last $|F_b|$ columns of $K_{-FF}$.
Similarly, the surviving lines in $F_{-1}$ correspond to the first $|F_{-1}|$ rows of $K_{-FF}$, and 
the surviving lines in $F_{-b}$ correspond to the last $|F_{-b}|$ rows of $K_{-FF}$.  
The ordering of rows and columns of $D_{-FF}$ is the same as that for $K_{-FF}$.
Similarly the rows and columns of $D_{FF}$ will be ordered according to $F_k$, $k=1, \dots, b$.
Finally partition $B$ and $C$ according to the block structures of both $-F, F$ and $\calE$:
\begin{IEEEeqnarray*}{rCl}
B  &=:&  \begin{bmatrix} B_{-F} & 0 \\ 0 & B_F \end{bmatrix}\\
 & =:& \begin{bmatrix}  \text{diag}(B_{-k}, k = 1, \dots, b)  & 0 \\ 
 0 &   \text{diag}(B_{k}, k = 1, \dots, b)  \end{bmatrix},\\
C & =:&  \begin{bmatrix} C_{-F} & C_{F} \end{bmatrix}  \\
& =:& 	\begin{bmatrix} C_{-1} & \cdots & C_{-b} & \ & C_{1} & \cdots & C_{b} \end{bmatrix}.
\end{IEEEeqnarray*}
Recall that $D_{-FF} = B_{-F}C_{-F}^T A C_F$ and
$D_{FF} = B_F C_F^T A C_F$.   
Corollary~\ref{ch:cf; corollary:blockpartition} then implies that the PTDF submatrices $D_{-FF}$ and $D_{FF}$ 
decompose into diagonal structures corresponding to the blocks of $\calG$:
\begin{subequations}
\begin{align}
D_{-FF}   =: \begin{bmatrix} 
	D_{-1}  & 0 & \dots & 0  \\  0 & D_{-2} & \dots & 0 \\ \vdots & \vdots & \ddots & \vdots \\ 0 & 0 & \dots & D_{-b} 
	\end{bmatrix}, \label{ch:cf; sec:localization; eq:K-FF.1a}
\end{align}
where $D_{-k} := B_{-k} C_{-k}^T A C_k, \  k = 1, \dots, b$. Moreover,
\begin{align}
D_{FF} \ =: \ \begin{bmatrix} 
	D_1 & 0 & \dots & 0  \\  0 & D_2 & \dots & 0 \\ \vdots & \vdots & \ddots & \vdots \\ 0 & 0 & \dots & D_b 
	\end{bmatrix}, \label{ch:cf; sec:localization; eq:K-FF.1b}
\end{align}
where $D_k \ := \ B_k C_k^T A C_k, \  k = 1, \dots, b $.
Here each $D_{-k}$ is $|F_{-k}|\times |F_k|$ and each $D_k$ is $|F_k|\times |F_k|$.  They involve lines only 
in block $\calE_k$ of $\calG$.
Since $K_{-FF} = D_{-FF} ( I - \text{diag}(D_{FF}) )^{-1}$, the LODF
submatrix $K_{-FF}$ has the same block diagonal structure as $D_{-FF}$:
\begin{align}
K_{-FF} =:  \begin{bmatrix} 
	K_1 & 0 & \dots & 0  \\  0 & K_2 & \dots & 0 \\ \vdots & \vdots & \ddots & \vdots \\ 0 & 0 & \dots & K_b 
	\end{bmatrix},
\end{align}
where $K_{k}:=D_{-k} \left( I - \text{diag}(D_k) \right)^{-1}$, $k = 1, \dots, b$ 
and $D_{-k}, D_k$ are given in \eqref{ch:cf; sec:localization; eq:K-FF.1a}, \eqref{ch:cf; sec:localization; eq:K-FF.1b}.
As for $D_{-k}$, each $K_k$ is $|F_{-k}|\times |F_k|$ and involves lines only in block $\calE_k$ of $\calG$.
The invertibility of $I - \text{diag}(D_k)$ follows from Corollary \ref{ch:lm; corollary:PTDF.2}.
\label{ch:cf; sec:localization; eq:K-FF.1}
\end{subequations}

Even if $K^F \neq K_{-FF}$ in general, the next result shows that the GLODF $K^F$ has the same block-diagonal structure as $K_{-FF}$. This implies that even though the impacts of multiple simultaneous line outages are correlated through the network topology, such correlations are present only within each block. In particular, the impacts of a non-cut outage are also localized within blocks that contain outaged lines. It is proved by substituting \eqref{ch:cf; sec:localization; eq:K-FF.1} into Theorem \ref{ch:cf; thm:LODF.3}.
\begin{thm}[Failure localization: non-cut set outage]
\label{ch:cf; thm:K^Fblock.1}
Suppose a non-cut set $F$ of lines trip simultaneously so that the surviving graph 
$(\calN, \calE\!\setminus\! F)$ remains connected.  For any surviving line $l = (i, j)$:
\begin{enumerate}
\item GLODF $K^F_{l \hat l} = 0$ if $l$ and $\hat l$ are in different blocks of $\calG$.
\item $K^F := K^F_{-FF}$ has a block diagonal structure:
\begin{subequations}
\begin{eqnarray}
K^F   =: \begin{bmatrix} 
	K^F_1 & 0 & \dots & 0  \\  0 & K^F_2 & \dots & 0 \\ \vdots & \vdots & \ddots & \vdots \\ 0 & 0 & \dots & K^F_b
	\end{bmatrix},
\label{ch:cf; sec:localization; eq:K^F.1a}
\end{eqnarray}
where for $k=1, \dots, b$ each $K^F_k$ is $|F_{-k}|\times |F_k|$ and involves lines only in block $\calE_k$ of $\calG$, given by:
\begin{IEEEeqnarray}{rCl}
K^F_k  & := & D_{-k} (I - D_k)^{-1} \\ 
& = & K_k \left(I - \diag(D_k) \right) \left( I-D_k \right)^{-1},
\label{ch:cf; sec:localization; eq:K^F.1b}
\end{IEEEeqnarray}
or in terms of $B, C$ and $A$:
\begin{align}
K^F_k = B_{-k} C_{-k}^T A C_k \left( I-B_k C_k^T A C_k \right)^{-1}.
\label{ch:cf; sec:localization; eq:K^F.1c}
\end{align}
\label{ch:cf; sec:localization; eq:K^F.1}
\end{subequations}
\end{enumerate}
\end{thm}
Again, since a bridge is a block, a non-cut outage does not impact the branch flow on any bridge. The invertibility of $I - D_k$ follows from Corollary~\ref{ch:lm; corollary:PTDF.2} and the block-diagonal structure of $D_{FF}$.
Theorem \ref{ch:cf; thm:K^Fblock.1} subsumes Corollary \ref{ch:cf; corollary:blockpartition} which corresponds to  the special case where $F= \{\hat l\}$. In that case $K^F = K^{\hat l}$ is a size $m-1$ column vector.  If $\hat l\in \calE_1$ then $D_{FF} = D_{\hat l \hat l}$ and
\begin{eqnarray*}
K^{\hat l}  &=&  \begin{bmatrix} K_1 \\ 0 \\ \vdots \\ 0 \end{bmatrix},
\end{eqnarray*}
with $K_1 := D_{-1} (1-D_{\hat l\hat l})^{-1}$ and $D_{-1}:= (D_{l\hat l}, l\neq \hat l, l\in \calE_1)$.

The ability to characterize in terms of the GLODF $K^F$ the localization of the impact of line outages within each block where outages occur is illustrated in the next example.
\begin{eg}
\label{ch:cf; sec:localization; eg:N-2outage}
Consider a non-cut set $F := \{l_1, l_2\}$ and the $N-2$ event where lines $l_1$ and $l_2$ trip simultaneously. The branch flow change on a surviving line $l\in -F$ is given in terms of the GLODF $K^F$ defined in 
\eqref{ch:cf; subsec:lodf; eq:defKlF} as:
\begin{align*}
\tilde f_{l} - f_{l} & \ = \ K^F_{l l_1} f_{l_1} \ + \ K^F_{l l_2} f_{l_2},
\end{align*}
where $K^F_{l \hat l}$ is the $(l, \hat l)$-th entry of $K^F$, $i=1, 2$.
There are two cases:
\begin{enumerate}
\item \emph{Lines $l_1, l_2$ are in the same block $\calE_k$}. Then
	\begin{align*}
	\tilde f_{l} - f_{l} & \ = \ \left\{ \begin{array}{lcl} 0 & & \text{if } l \not\in \calE_k, \\
					K^F_{l l_1} f_{l_1} + K^F_{l l_2} f_{l_2} & & \text{if } l \in \calE_k.
					\end{array} \right.
	\end{align*}

\item \emph{Lines  $l_1, l_2$ are in different blocks, say $l_i\in \calE_i$}.
	Then
	\begin{align*}
	\tilde f_{l} - f_{l} & \ = \ \left\{ \begin{array}{lcl} 0 & & \text{if } l \not\in \calE_1 \cup \calE_2,\\
					K^F_{l l_1} f_{l_1} & & \text{if } l \in \calE_1, \\
					K^F_{l l_2} f_{l_2} & & \text{if }l \in \calE_2.
					\end{array} \right.
	\end{align*}
	In this case since there is a single non-bridge line that is outaged in each block, the decoupling of
	outages over different blocks means
	$K^F_{l \hat l} = K_{l \hat l}$ as if each of the outaged lines $l_1$ and $l_2$ is outaged
	separately.
\end{enumerate}
\qed
\end{eg}

Theorem~\ref{ch:cf; thm:K^Fblock.1} is a consequence of the Simple Cycle Criterion since $K_{l \hat l} \neq 0$ only if there is a simple cycle that contains both $l$ and $\hat l$.
The converse of the Simple Cycle Criterion asserts that $K_{l \hat l} \neq 0$ ``if'' there is a simple cycle that contains both $l$ and $\hat l$.  This immediately implies the converse of Corollary \ref{ch:cf; corollary:blockpartition} that $K_{l \hat l} \neq 0$ ``if" $l$ and $\hat l$ are in the same block of $\calG$.  In other words, not only the submatrices $K_{-FF},  D_{-FF}, D_{FF}$ are block-diagonal, but also that almost surely with respect to $\mu$, \emph{every} entry of the diagonal blocks $K_k, D_{-k}, D_k$ in \eqref{ch:cf; sec:localization; eq:K-FF.1} is nonzero.
This is only for the case when a single non-bridge line $\hat l$ trips. It does not directly imply the converse of Theorem \ref{ch:cf; thm:K^Fblock.1}, i.e., it is not clear whether every entry of $K^F_k$ is nonzero ($\mu$-almost surely) when multiple lines in a non-cut set $F$ trip simultaneously. Even though every entry of $D_{-k}, D_k$ is nonzero ($\mu$-almost surely), the issue is whether every entry of the product $K^F_k = D_{-k} (I - D_k)^{-1}$ from \eqref{ch:cf; sec:localization; eq:K^F.1b} is nonzero ($\mu$-almost surely).
The next result (proved in Appendix \ref{proof:them:K^Fblock.2}) shows indeed the converse of Theorem \ref{ch:cf; thm:K^Fblock.1} holds as well.
\begin{thm}[Failure localization: converse]
\label{ch:cf; thm:K^Fblock.2}
Suppose a non-cut set $F$ of lines trip simultaneously so that the surviving graph 
$(\calN, \calE\!\setminus\! F)$ remains connected.  
\begin{enumerate}
\item For any surviving line $l= ( i, j)$ the GLODF $K^F_{l \hat l} \neq 0$ ``if'' and only if $l$ and $\hat l$ 
	are in the same block of $\calG$.
\item Every entry of the diagonal blocks $K^F_k, K_k, D_{-k}, D_k$ in 
	\eqref{ch:cf; sec:localization; eq:K^F.1} and \eqref{ch:cf; sec:localization; eq:K-FF.1} is nonzero
	$\mu$-almost surely.
\end{enumerate}
\end{thm}

%% file: conclusion.tex
In Part I of this work, we develop a spectral theory using the transmission network Laplacian matrix that precisely captures the Kirkhhoff's Law in terms of graphical structures. Our results show that the distributions of different families of subtrees play an important role in understanding power redistribution and enables us to derive algebraic properties using purely graphical arguments. We consider the scenario where the surviving network remains connected and explain how the localizability of line failures can be fully characterized using its block decomposition.


%% file: proofs.tex
\section{Appendix: proofs}
\label{section:proofs}

\subsection{Proof of Theorem \ref{prop:prem_basis}}\label{section:spectral_rep_proof}
Recall that, without loss of generality, we choose node $n$ as reference node and defined the matrix $A$ accordingly in \eqref{eqn:extension_of_a}. If $i=n$ or $j=n$, it is easy to see that $\calT(ij, n)=\emptyset$ so that $A_{ij}=0$. Now suppose $i,j\neq n$, we can express $A_{ij}$ in terms of the weighted spanning trees through the Cramer's rule. Specifically, let $A_j$ denote the $j$-th column of $A$ after removing the reference node. Note from the definition of $A$ that $\ol{L}A_{j}=e_j$,
where $e_j\in\R^{n-1}$ is the vector with $1$ as its $j$-th component and $0$ elsewhere. Cramer's rule gives
$$
A_{ij}=\frac{\det\paren{\ol{L}^i_j}}{\det\paren{\ol{L}}},
$$
where $\ol{L}^i_j$ is the matrix obtained by replacing the $i$-th column of $\ol{L}$ by $e_j$. Now, by Lemma~\ref{prop:prem_laplace_tree}, we have
$$
\det\paren{\ol{L}^i_j}=(-1)^{i+j}\det\paren{\ol{L}^{ij}}=\sum_{F\in \calT(ij, n)}\beta(F),
$$
and using the Kirchhoff's Matrix Tree Theorem we obtain
$$
\det\paren{\ol{L}}=\sum_{F\in \calT_\calE}\beta(F),
$$
concluding the proof. \hfill \qed

\subsection{Proof of Theorem \ref{cor:prem_cross_product}}\label{section:prem_proof_of_cross}
Theorem \ref{prop:prem_basis} implies that
\begin{IEEEeqnarray}{rCl}
&\paren{\sum_{F\in \calT_\calE}\beta(F)}\paren{A_{i\hat i} + A_{j\hat j} - A_{i\hat j} - A_{j\hat i}} \nonumber\\
&=\sum_{F\in \calT(i \hat i,n)}\beta(F) + \sum_{F\in \calT(j \hat j,n)}\beta(F)\nonumber \\
&-\sum_{F\in \calT(i \hat j,n)}\beta(F) - \sum_{F\in \calT(j \hat i,n)}\beta(F).\label{eqn:prem_decomposed1}
\end{IEEEeqnarray}
We can decompose the set $\calT(i\hat i, n)$ based on the tree to which node $j$ belongs. This leads to the identity
$$
\calT(i \hat i, n)=\calT(ij \hat i,n)\sqcup \calT(i\hat i,jn),
$$
where $\sqcup$ denotes a disjoint union. Similarly, we also have
$$
\calT(j\hat j,n)=\calT(ij \hat j ,n)\sqcup \calT(j\hat j,in),
$$
$$
\calT(i \hat j, n)=\calT(ij \hat j, n)\sqcup \calT(i \hat j, jn),
$$
$$
\calT(j \hat i, n)=\calT(i j \hat i, n)\sqcup \calT(j \hat i, in).
$$
Substituting the above decompositions into \eqref{eqn:prem_decomposed1} and simplifying, we obtain
\begin{IEEEeqnarray}{rCl}
&\paren{\sum_{F\in \calT_\calE}\beta(F)}\paren{A_{i\hat i} + A_{j\hat j} - A_{i\hat j} - A_{j\hat i}} \nonumber\\
&\quad =\sum_{F\in \calT(i\hat i, jn)}\beta(F) + \sum_{F\in \calT(j \hat j, in)}\beta(F)\nonumber\\
&\quad -\sum_{F\in \calT(i \hat j, jn)}\beta(F) - \sum_{F\in \calT(j\hat i, in)}\beta(F).\label{eqn:prem_decomposed2}
\end{IEEEeqnarray}
Furthermore, the following set of identities hold:
\begin{IEEEeqnarray*}{rCl}
\calT(i\hat i, jn)&=&\calT(i \hat i, jn \hat j)\sqcup\calT(i \hat i \hat j, jn),\\
\calT(j \hat j, in)&=&\calT(j\hat j, in \hat i)\sqcup\calT(j \hat i \hat j, in),\\
\calT(j \hat i, in)&=&\calT(j\hat i, i n \hat j)\sqcup\calT(j \hat i \hat j,in),\\
\calT(i\hat j, jn)&=&\calT(i\hat j, jn \hat i)\sqcup\calT(i \hat i \hat j, jn).
\end{IEEEeqnarray*}
Substituting these into \eqref{eqn:prem_decomposed2} and rearranging yields
\begin{IEEEeqnarray*}{rCl}
&\paren{\sum_{F\in \calT_\calE}\beta(F)}\paren{A_{i\hat i} + A_{j\hat j} - A_{i\hat j} - A_{j\hat i}}\\
&\quad=\sum_{F\in \calT(i \hat i, j n \hat j)}\beta(F) + \sum_{F\in \calT(j \hat j, in \hat i)}\beta(F) \\
&\quad-\sum_{F\in \calT(j \hat i, in \hat j)}\beta(F) - \sum_{F\in \calT(i \hat j, jn \hat i)}\beta(F)\\
&\quad=\sum_{F\in \calT(i \hat i, j \hat j)}\beta(F) - \sum_{F\in \calT(j \hat i, i \hat j)}\beta(F),
\end{IEEEeqnarray*}
where the last equality follows from
\begin{IEEEeqnarray*}{rCl}
\calT(i \hat i, j\hat j)&=&\calT(i \hat i, jn \hat j)\sqcup \calT(j\hat j, in \hat i)
\end{IEEEeqnarray*}
and
\begin{IEEEeqnarray*}{rCl}
\calT(j \hat i, i \hat j)&=&\calT(j \hat i, in \hat j)\sqcup \calT(i \hat j, jn \hat i).
\end{IEEEeqnarray*}
This completes the proof.
\qed

\subsection{Proof of Theorem \ref{ch:cf; thm:LODF.3}}\label{appendix:proof_of_glodf}
The first part is proved by analyzing the post-contingency network $A_{-F}$, the second part is proved by analyzing the pre-contingency graph $A$, and the third part is proved by relating $D_{-FF}$ and $K_{-FF}$.
\paragraph{Proof based on post-contingency $A_{-F}$}
The DC power flow equations \eqref{eqn:dc_model} for the pre-contingency network are:
\begin{equation}
    p = Cf = C_{-F}f_{-F} + C_{F}f_F, \quad f = BC^T \theta. \label{eqn:thm8.pre-DC}
\end{equation}
Let $(\tilde{f}_{-F}, \tilde{\theta})$ denote the post-contingency branch flows and phase angles. Given the assumption that the power injections remain the same, we have the following DC power flow equations for the post-contingency network:
\begin{equation}
    p = C_{-F}\tilde{f}_{-F}, \quad \tilde{f}_F = B_{-F}C_{-F}^T \tilde{\theta}. \label{eqn:thm8.post-DC}
\end{equation}
Subtracting \eqref{eqn:thm8.pre-DC} from \eqref{eqn:thm8.post-DC} gives
\begin{equation*}
    C_{-F} (\tilde{f}_{-F} - f_{-F}) = C_Ff_F, \quad \tilde{f}_{-F} - f_{-F} = B_{-F}C_{-F}^T (\tilde{\theta} - \theta).
\end{equation*}
Therefore, $(\Delta f_{-F}, \Delta \theta) := (\tilde{f}_{-F} - f_{-F}, \tilde{\theta} - \theta)$ satisfies the DC power flow equations with injections $C_Ff_F$ on the post-contingency network. The unique solution for $\Delta f_{-F}$ is:
\begin{equation*}
    \Delta {f}_{-F} = \underbrace{B_{-F}C_{-F}^TA_{-F}C_F}_{K^F} f_F.
\end{equation*}
\paragraph{Proof based on pre-contingency $A$}
In this part, we construct a fictitious network that mimics the impact of the non-cut $F$ outage. Specifically, the network is the same as the pre-contingency netowrk but with its injections changed from $p$ to $\hat p = p+\Delta p$. For this fictitious network, the DC power flow equations write:
\begin{equation}
    \hat p = C \hat f = C_{-F} \hat f_{-F} + C_{F} \hat f_{F}, \quad \hat f = BC^T\hat \theta. \label{eqn:fic_dc}
\end{equation}
We choose $\Delta p$ so that it is carried entirely by the fictitious branch flows $\hat f_{F}$ on lines in $F$ that would be have been disconnected, i.e. we pick
\begin{equation}
    \Delta p = C_{F} \hat f_{F} \label{eqn:thm8.fic_inj}.
\end{equation}
This additional injection $\Delta p$ does satisfy $\bm{1}^T \Delta p =0$ and is thus balanced. Substituting \eqref{eqn:thm8.fic_inj} into \eqref{eqn:fic_dc} yields:
\begin{equation}
    p = C_{-F} \hat f_{-F}, \quad \hat f_{-F} = B_{-F} C_{-F}^T \hat \theta,
\end{equation}
i.e., $(\hat f_{-F},\hat \theta)$ satisfies the same DC power flow equations \eqref{eqn:thm8.post-DC} for the post-contingency network. Since the DC power flow equations have an unique branch flow solution, the post-contingency branch flows $\tilde{f}_{-F}$ from \eqref{eqn:thm8.post-DC} must coincide with the branch flows $\hat f_{-F}$ in the fictitious network \eqref{eqn:fic_dc}.
This allows us to calculate the GLODF $K^F$ by relating $\tilde{f}_{-F}$ and $f_{F}$ on two different networks through the relation between $\hat f_{-F}$ and $f_{F}$ on the same pre-contingency network. Considering the fictitious network, we have:
\begin{equation*}
    \hat f_{F} = B_F C_F^T \hat \theta = B_F C_F^T A (p+\Delta p)
\end{equation*}
Substituting into \eqref{eqn:thm8.fic_inj} gives $\Delta p = C_F \hat f_F = C_F B_F C_F^T A (p+\Delta p)$. Hence,
\begin{equation}
    \Delta p = \left(I - C_FB_FC_F^T A \right)^{-1} C_FB_FC_F^TAp,
\end{equation}
which yields the following expression for $\hat f_{-F}$ in the fictitious network
\begin{equation*}
    \hat f_{-F} = B_{-F}C_{-F}^T A (p+\Delta p).
\end{equation*}
The pre-contingency line flows are given by
\begin{equation*}
    f_{-F} = B_{-F}C_{-F}^T A p, \quad f_{F} = B_{F}C_{F}^T A p.
\end{equation*}
Substituting these expressions we have
\begin{IEEEeqnarray}{rCl}
\hat f_{-F} &=& f_{-F} + B_{-F}C_{-F}^T A \left(I - C_FB_FC_F^T A \right)^{-1} C_F f_F \nonumber \\
& = & f_{-F} + \underbrace{B_{-F}C_{-F}^T A C_F \left(I - B_FC_F^T A C_F\right)^{-1}}_{K^F} f_F, \nonumber
\end{IEEEeqnarray}
where we use the identity $(I+M_1M_2)^{-1}M_1 = M_1(I+M_2M_1)^{-1}$ (provided the inverse exists) in the last equality. This identity follows from:
\begin{eqnarray*}
M_1 &=&M_1(I+M_2M_1)(I+M_2M_1)^{-1} \\
&=&(M_1+M_1M_2M_1)(I+M_2M_1)^{-1} \\
&=& (I+M_1M_2)M_1(I+M_2M_1)^{-1}.
\end{eqnarray*}
\paragraph{Relation between $K^F$ and $K_{-FF}$}
As shown in \eqref{eqn:K_-FF}, we have $D_{-FF}=K_{-FF}(I-\text{diag}(D_FF))$. The $K^F$ in terms of pre-contingency network $A$ yields:
\begin{eqnarray*}
K^F &=& B_{-F}C_{-F}^T A C_F \left(I - B_FC_F^T A C_F\right)^{-1} \\
& = & D_{-FF} \left( I - D_{FF}\right)^{-1} \\
& = & K_{-FF}(I-\text{diag}(D_FF))\left( I - D_{FF}\right)^{-1}. \hfill \qed
\end{eqnarray*}

\subsection{Proof of Theorem \ref{prop:simple_loop}}
\label{proof:SimpleLoopC}
Theorem \ref{thm:redist_factor} implies that $K_{l \hat l}$ is proportional to 
the following polynomial in the subsceptances $B$:
\begin{eqnarray*}
f (B) & := & \sum_{F\in \calT(i \hat i, j \hat j)}\beta(F) - \sum_{F\in \calT(j \hat i, i \hat j)}\beta(F).
\end{eqnarray*}
If $K_{l \hat l} \neq 0$ then at least one of the sets $\calT(i \hat i, j\hat j)$ and $\calT(i \hat j, j \hat i)$ of spanning forests is nonempty.  
Suppose $\calT(i\hat i, j\hat j)$ is nonempty and contains a spanning forest $F$.  
The tree in $F$ that contains buses $i$ and $\hat i$ defines a path from $i$ to $\hat i$, and the other tree that 
contains $j$ and $\hat j$ defines a path from $j$ to $\hat j$.  These two paths are vertex-disjoint, i.e., they do not share any 
vertices.  If we add the lines $l = (i, j)$ and $\hat l = (\hat i, \hat j)$ to these two vertex-disjoint paths we obtain a simple cycle 
that contains both $l$ and $\hat l$.

Conversely suppose there is a simple cycle that contains $l$ and $\hat l$.  
Removing lines $l$ and $\hat l$ from the simple cycle
produces two vertex-disjoint paths, say, $P_i$ that contains buses $i, \hat i$ and $P_j$ that
contains buses $j, \hat j$.  Since $\calG$ is connected we can extend $P_i$ and $P_j$ into a spanning forest with exactly
two disjoint trees.  This spanning forest, denoted by $F$, is in $\calT(i\hat i, j\hat j)$.  
Furthermore $F$ is not in $\calT(i\hat j, j \hat i)$ from the following claim:  $$\calT(i\hat i, j \hat j) \cap \calT(i\hat j, j \hat i) = \emptyset.$$
To show this, consider an element $T_1$ from $\calT(i\hat i, j\hat j)$, which consists of two trees $\calT_1$ and $\calT_2$ with $\calT_1$ containing $i,\hat i$ and $\calT_2$ containing $j , \hat j$. If $T_1\in\calT(i \hat j, j \hat i)$, then $\calT_1$ must also contain $\hat j$. However, this implies $\hat j\in \calT_1\cap \calT_2$, and thus $\calT_1$ and $\calT_2$ are not disjoint, contradicting the definition of $\calT(i \hat j, j \hat i)$.
Hence $f(B)$ is not identically zero.
This means $K_{l \hat l} = 0$ if and only if $B$ is a root of the polynomial $f(B)$.  
It is a fundamental result that the root set of a polynomial which is not identically zero has Lebesgue measure
zero.  Therefore, since $\mu$ is absolutely continuous with respect to the Lebesgue measure ${\cal L}_m$, we have
\begin{eqnarray*}
\mu(f(B+\omega) = 0) =  {\cal L}_m(f(B+\omega) = 0)  = 0
\end{eqnarray*}
i.e., $\mu(K_{l \hat l}\neq 0) = \mu(f(B+\omega) \neq 0)  = 1$ if there is a simple cycle that contains $l$ and $\hat l$. \qed

\subsection{Proof of Theorem \ref{ch:cf; thm:K^Fblock.2}}\label{proof:them:K^Fblock.2}
We only provide a sketch of the proof (see \cite{thesis} for details). 
 It suffices to show that if $l$ and $\hat l$ are in the same block of $\calG$ then $K^F_{l\hat l} \neq 0$
$\mu$-almost surely.

Recall that $K^F_k = D_{-k} (I - D_k)^{-1}$, where the $(l, \hat l)$-th entry of $D_{-k}$ or $D_k$ is given by
\begin{equation}
D_{l, \hat{l}} = \frac{B_l \left(\sum_{F\in \calT(i\hat i, j \hat j)}\beta(F) -\sum_{F\in \calT(i \hat j, j \hat i)}\beta(F) \right)}{\sum_{F\in \calT_\calE}\beta(F)},
\label{ch:cf; sec:localization; eq:K^Fconverse.1}
\end{equation}
where $l = (i,j)$ and $\hat l = (\hat i, \hat j)$. By construction, $K^F, D_{-k}, D_k$ involve lines only in block $\calE_k$ of $\calG$. Consider
\begin{equation}
K^F_{l\hat l} = \sum_{m\in \calE_k} \left[ D_{-k} \right]_{lm} \left[ \left( I - D_k \right)^{-1} \right]_{m\hat l}, \label{ch:cf; sec:localization; eq:K^Fconverse.2}
\end{equation}
where $[M]_{l\hat l}$ denotes the $(l, \hat l)$-th entry of a matrix $M$.
We treat $D_{l\hat l} := D_{l\hat l}(B)$ and hence $K^F_{l\hat l} := K^F_{l\hat l}(B)$ for each pair 
$(l, \hat l)\in \calE_1\times \calE_1$ as polynomials in the susceptances $B = (b_l, l\in \calE)$. Without loss of generality we focus on the first block $\calE_1$.   
The proof consists of three steps: (1) show that $D_{l\hat l}(B)$ for each $(l, \hat l)\in \calE_1\times \calE_1$ is not a zero polynomial; (2) show that $[ \left( I - D_k(B) \right)^{-1}]_{l\hat l}$ for each $(l, \hat l)\in \calE_1\times \calE_1$ is not a zero polynomial; and (3) show that the summands in \eqref{ch:cf; sec:localization; eq:K^Fconverse.1} do not cancel out so that $K^F_{l\hat l}(B)$ is not a zero polynomial.  Then $\mu(K^F_{l\hat l}(B + \omega) = 0) = 0$, i.e., $K^F_{l\hat l} \neq 0$ $\mu$-almost surely.

Step 1 can be proved by applying the same argument in the proof of the Simple Cycle Criterion (Theorem 8 in Part I) for $K_{l\hat l}$ to the expression \eqref{ch:cf; sec:localization; eq:K^Fconverse.1} of $D_{l\hat l}$. For Step 2 let $M := (I - D_k)^{-1}$.  Then column $\hat l$ of $(I - D_k) M = I$ gives $(I - D_k) M_{\hat l} = e_{\hat l}$ where $M_{\hat l}$ is column $\hat l$ of $M$ and $e_{\hat l}$ is the standard unit vector. Cramer's rule then implies
\begin{IEEEeqnarray*}{rCl}
M_{l\hat l} & = & \left[ \left( I - D_k \right)^{-1} \right]_{l\hat l} \ \ = \ \ 
		\frac{ (-1)^{l + \hat l} \ \text{det} \left( I - D_k^{\hat l l} \right) } { \text{det} (I - D_k) },
\end{IEEEeqnarray*}
where $D_k^{\hat l l}$ is obtained from the $|F_k| \times |F_k|$ matrix $D_k$ by removing its row $\hat l$ and column $l$ and $I$ above is the size $|F_k|-1$ identity matrix. Leibniz's formula for determinant then yields
\begin{IEEEeqnarray*}{rCl}
\text{det} \left( I - D_k^{\hat l l}(B) \right) & = & 
		\sum_\sigma \text{sgn}(\sigma) \prod_m \left( I - \left[ D_k^{\hat l l} \right]_{m\sigma(m)}(B) \right).
\end{IEEEeqnarray*}
It is proved in \cite{thesis} that the determinant above is not a zero polynomial in $B$ for each pair of 
$(l, \hat l)\in \calE_1\times \calE_1$, including $l = \hat l$.
This implies $M_{l\hat l} := M_{l \hat l}(B)$ is not zero polynomials in $B$.
Lastly, step 3 is proved in \cite{thesis} and hence $K^F_{l\hat l}(B)$ is not a zero polynomial in $B$.
This completes the proof sketch. \qed

%% file: bios.tex
\begin{IEEEbiography}[{\includegraphics[width=1in,height=1.25in,clip,keepaspectratio]{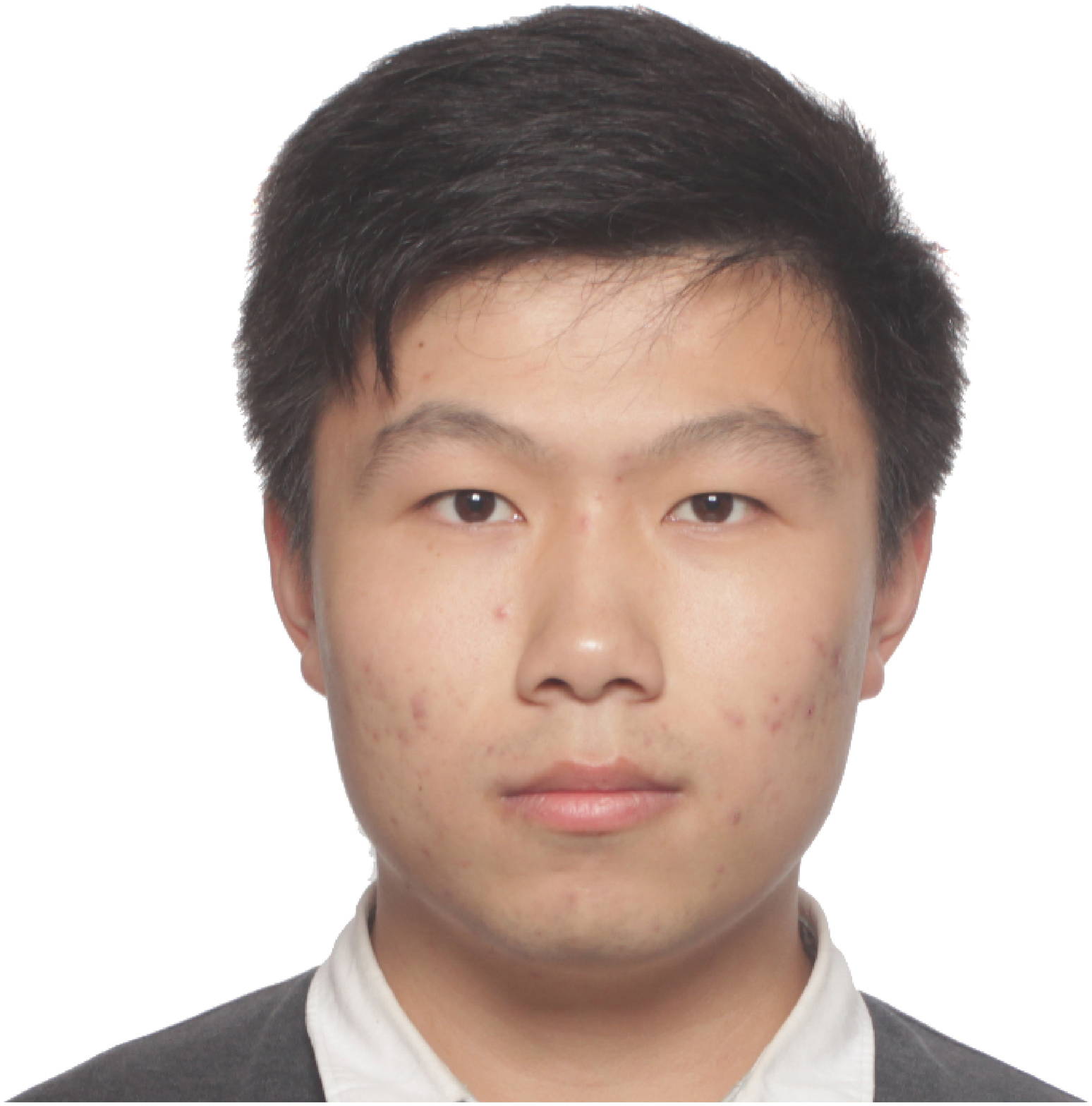}}]{Linqi Guo}
 received his B.Sc.~in Mathematics and B.Eng.~in Information Engineering from The Chinese University of Hong Kong in 2014, and his Ph.D.~in Computing and Mathematical Sciences from California Institute of Technology in 2019. His research is on the control and optimization of networked systems, with focus on power system frequency regulation, cyber-physical network design, distributed load-side control, synthetic state estimation and cascading failure analysis.
\end{IEEEbiography}

\begin{IEEEbiography}[{\includegraphics[width=1in,height=1.25in,clip,keepaspectratio]{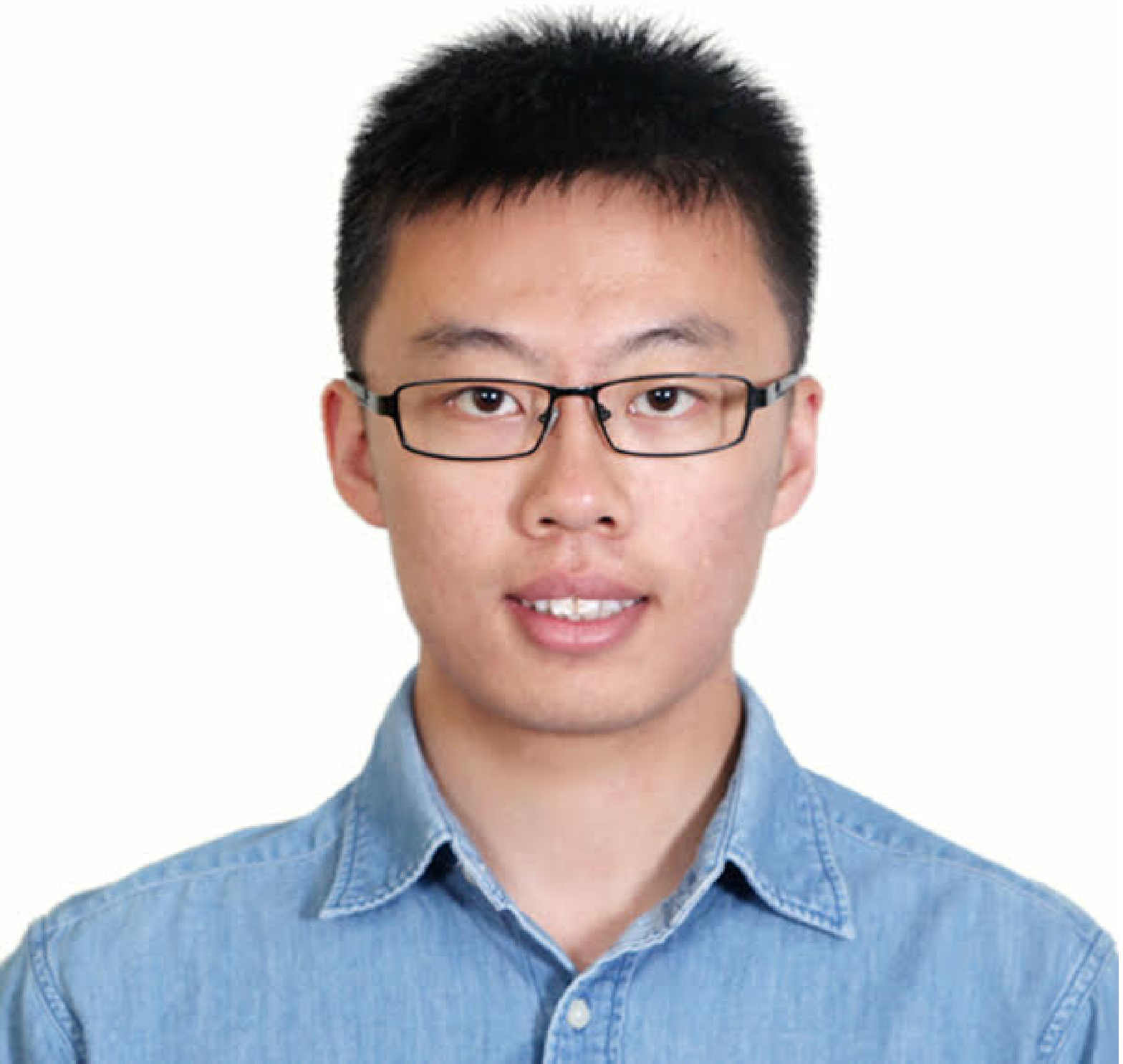}}]{Chen Liang (S’19)}
 received the B.E. degree in automation from Tsinghua University, Beijing, China, in 2016. He is currently pursuing the Ph.D. degree in Computing and Mathematical Sciences with the California Institute of Technology, Pasadena, CA, USA. 
His research interests include graph theory, mathematical optimization, control theory, and their applications in cascading failures of power systems.
\end{IEEEbiography}

\begin{IEEEbiography}[{\includegraphics[width=1in,height=1.25in,clip,keepaspectratio]{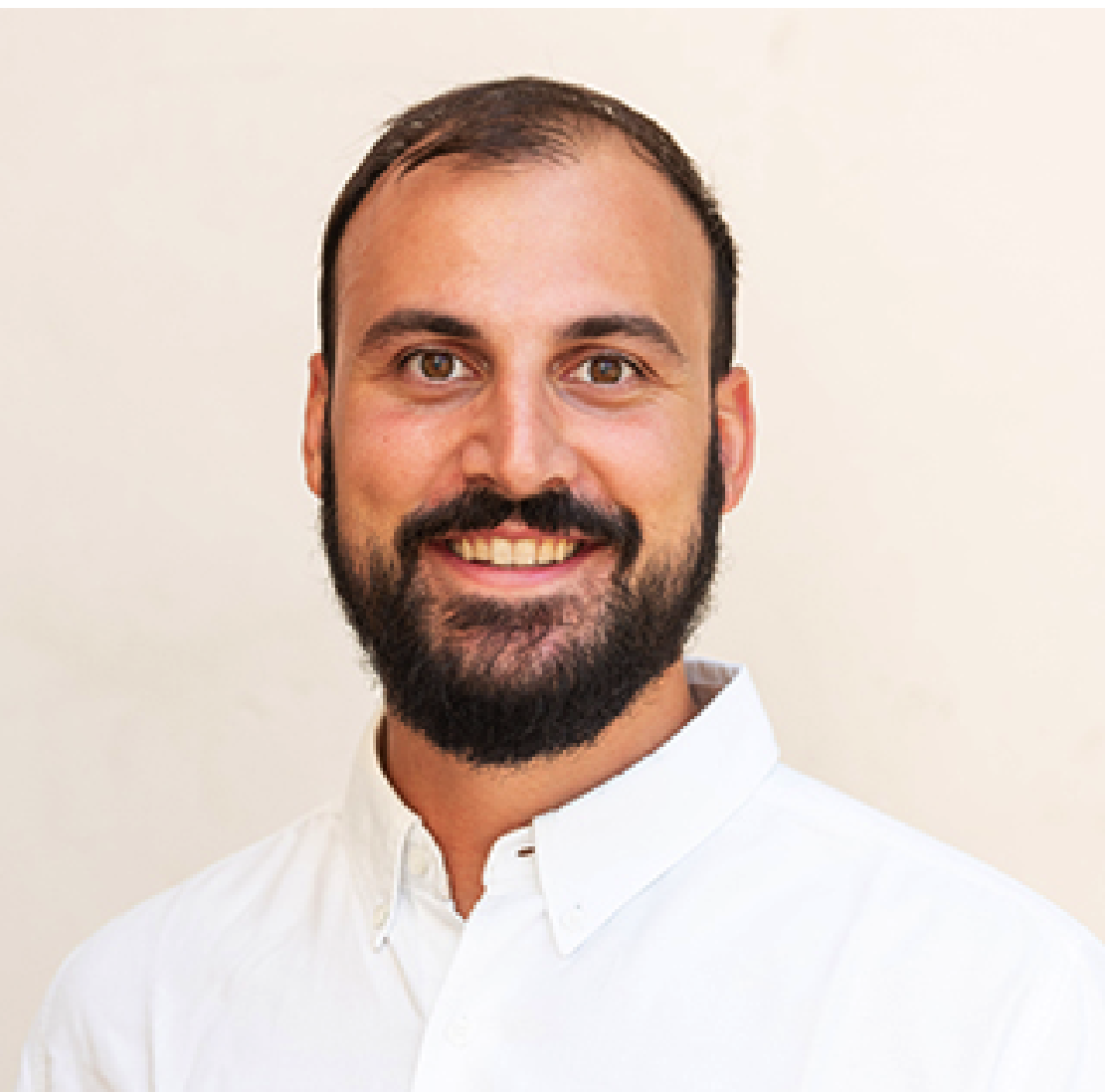}}]{Alessandro Zocca}
 received his B.Sc. in mathematics from the University of Padua, Italy, in 2010, his M.A.St. in mathematics from the University of Cambridge, UK, in 2011, and his Ph.D. degree in mathematics from the University of Eindhoven, The Netherlands, in 2015. He then worked as postdoctoral researcher first at CWI Amsterdam (2016-2017) and then at California Institute of Technology (2017-2019), where he was supported by his personal NWO Rubicon grant. Since October 2019, he has a tenure-track assistant professor position in the Department of Mathematics at the Vrije Universiteit Amsterdam. His work lies mostly in the area of applied probability and optimization, but has deep ramifications in areas as diverse as operations research, graph theory, algorithm design, statistical physics, and control theory. His research focuses on dynamics and rare events on large-scale networked systems affected by uncertainty, drawing motivation from applications to power systems and wireless networks.
\end{IEEEbiography}

\begin{IEEEbiography}[{\includegraphics[width=1in,height=1.25in,clip,keepaspectratio]{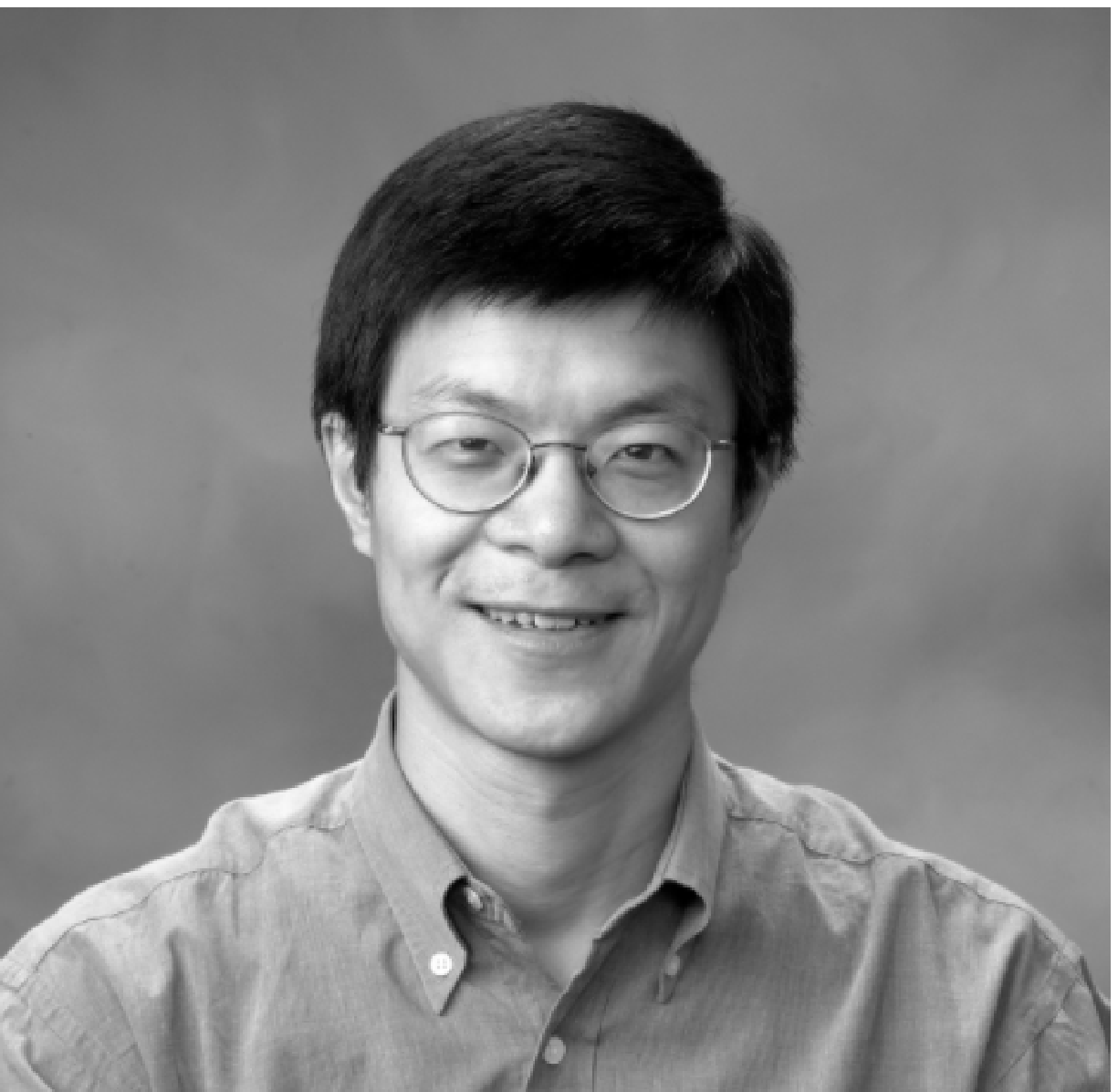}}]{Steven H. Low (F’08)}
is the F.~J.~Gilloon Professor of the Department of Computing \& Mathematical Sciences and the Department of Electrical Engineering at Caltech. Before that, he was with AT\&T Bell Laboratories, Murray Hill, NJ, and the University of Melbourne, Australia. He has
held honorary/chaired professorship in Australia, China and Taiwan. He was a co-recipient of IEEE best paper awards and is a Fellow of both IEEE and ACM. He was known for pioneering a mathematical theory of Internet congestion control and semidefinite relaxations of optimal power flow problems in smart grid.  He received his B.S. from Cornell and PhD from Berkeley, both in electrical engineering.
\end{IEEEbiography}

\begin{IEEEbiography}[{\includegraphics[width=1in,height=1.25in,clip,keepaspectratio]{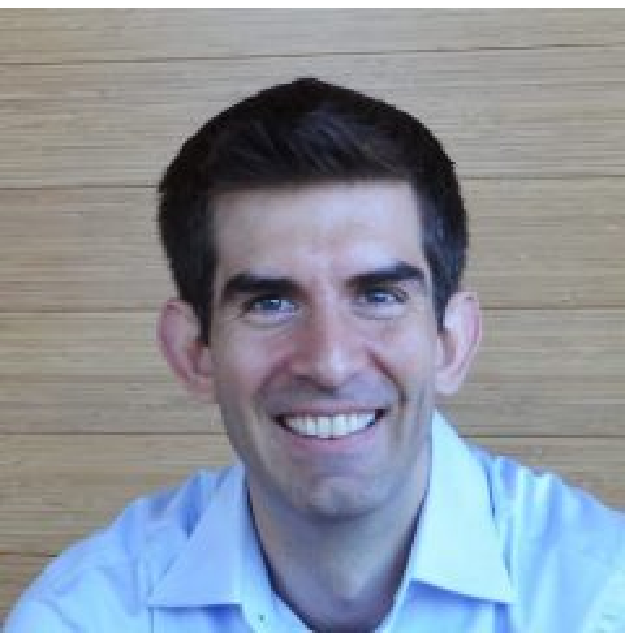}}]{Adam Wierman}
is a Professor in the Department of Computing and Mathematical Sciences at the California Institute of Technology. He received his Ph.D., M.Sc.~and B.Sc.~in Computer Science from Carnegie Mellon University in 2007, 2004, and 2001, respectively, and has been a faculty at Caltech since 2007. He is a recipient of multiple awards, including the ACM SIGMETRICS Rising Star award, the IEEE Communications Society William R. Bennett Prize, and multiple teaching awards. He is a co-author of papers that have received best paper awards at a wide variety of conferences across computer science, power engineering, and operations research including ACM Sigmetrics, IEEE INFOCOM, IFIP Performance, and IEEE PES.
\end{IEEEbiography}